\documentclass[11pt]{article}%
\usepackage{amsmath}
\usepackage{graphicx}
\usepackage{float}
\usepackage{caption}
\usepackage{subfigure}
\usepackage{array}
\usepackage{booktabs}%
\usepackage{amsfonts}%
\usepackage{amssymb}
\newcommand{\eqnb}{\begin{equation}}
\newcommand{\eqne}{\end{equation}}

\newtheorem{The}{Theorem}

\newtheorem{Def}{Definition}

\newtheorem{Lem}{Lemma}

\begin{document}

\title{Tree Representation, Growth Rate of Blockchain and Reward Allocation in
Ethereum with Multiple Mining Pools}
\author{Quan-Lin Li, Yan-Xia Chang, Chi Zhang\\School of Economics and Management \\Beijing University of Technology, Beijing 100124, China}
\maketitle

\begin{abstract}
It is interesting but difficult and challenging to study Ethereum with
multiple mining pools. One of the main difficulties comes from not only how to
represent such a general tree with multiple block branches (or sub-chains)
related to the multiple mining pools, but also how to analyze a
multi-dimensional stochastic system due to the mining competition among the
multiple mining pools. In this paper, we first set up a mathematical
representation for the tree with multiple block branches. Then we provide a
block classification of Ethereum: Regular blocks (in the main chain), orphan
blocks, uncle blocks, stale blocks, and nephew blocks, and give some key
probabilities of generating the different types of blocks by applying the law
of large numbers. Based on this, we further discuss the growth rate of
blockchain, and the reward allocation among the multiple mining pools through
applying the renewal reward theorem. Finally, we use some simulation
experiments to verify our theoretical results, and show
that the approximate computation approaches developed, such as the key probabilities,
the long-term growth rate of blockchain, and the long-term reward allocation
(rate) among the multiple mining pools, can have a faster convergence.
Therefore, we provide a powerful tool for observing and understanding the
influence of the selfish mining attacks on the performance of Ethereum with
multiple mining pools. We believe that the methodology and results developed in
this paper will shed light on the study of Ethereum with multiple mining pools,
such that a series of promising research can be inspired potentially.

\vskip                                                  0.5cm


\end{abstract}

\section{Introduction}

\label{Sec-1:introduction}

Bitcoin and blockchain have opened a new era of automatically processing and
storing transactions since the pioneering work by Nakamoto
\cite{Nak:2008}. The transactions can be automatically processed in the form
of Bitcoin script through a language of Merkle tree in a P2P network. However,
the Bitcoin language is not Turing-complete owing to robust concerns. On the other hand,
Ethereum can break such a limitation by introducing Ethereum Virtual Machine
featuring smart contract functionality, e.g., see \cite{But:2013, But:2014}.
Note that Bitcoin and Ethereum are the two largest and most popular
blockchain-based cryptocurrencies in the world. Ethereum has gained
great attention in the development of blockchain technology. Interested readers are referred
to Ethereum survey papers, for example, Wood \cite{Wood:2017}, Vuji\v{c}i\'{c}
et al. \cite{Vuj:2018}, Di Angelo et al. \cite{Di:2019}, Mohammed et al.
\cite{Moh:2021}; Ethereum system security by Chen et al. \cite{ChenP:2020} and
Praitheeshan et al. \cite{Pra:2019}; and smart contracts by Wang et al.
\cite{WangY:2018}, Dika and Nowostawski \cite{Dika:2018}, Wang et al.
\cite{WangJ:2021} and Atzei et al. \cite{Atzei:2017}. At the same time,
Ethereum has been developed into many practical applications from industry and
academia, e.g., see sharing economy by Bogner et al. \cite{Bogner:2016},
healthcare by Sookhak et al. \cite{Sook:2021}, IoT and logistics by Augusto et
al. \cite{Augusto:2019}, emergency service by Aung and Tantidham
\cite{Aung:2019}, decentralized marketplace by Ranganthan et al.
\cite{Ranganthan:2018}, and so forth.

Bitcoin and Ethereum applied the most widely used consensus mechanism:
Proof-of-Work (PoW). In a PoW blockchain system, many miners (or mining pools)
competitively mine each block which is generated by means of finding a nonce
through solving a cryptographic puzzle of using all the foregoing information
of that blockchain in front of this block, and then peg the block with the
nonce to the blockchain. See Li et al. \cite{Li:2018, Li:2021} for more
details. For the PoW blockchain system with two mining pools (honest and
dishonest), Eyal and Sirer \cite{Eya:2014} found the selfish mining attacks
and applied a simple Markov chain to study a few advantages of selfish mining.
Li et al. \cite{Li:2021} proposed the two-block leading competitive criterion,
and set up a pyramid Markov (reward) process to analyze the operations
efficiency and economic benefit of the blockchain selfish mining system. It is
seen from \cite{Eya:2014} and \cite{Li:2021} that the mining competition
between the two mining pools can be described as a comparison of lengths of
the two block branches, and the dishonest mining pool develops a selfish
mining attack policy, see Figures 2 and 4 in Li et al. \cite{Li:2021}.
Clearly, the longest one of the two block branches is called the main chain,
which is pegged on the blockchain; while another block branch is regarded as
the chain of orphan blocks, which cannot be connected to blockchain and is
returned to the transaction pool for reprocessing. Thus the orphan blocks
generate a lot of waste of computing resources. Li et al. \cite{Li:2021}
applied the pyramid Markov (reward) process to give a detailed analysis for
the orphan blocks and their waste of computing resources. Following the two
block branches corresponding to the two mining pools, this paper analyzes the
PoW blockchain system with multiple mining pools, and finds a general tree
with multiple block branches where the dishonest block branches can fork at
different positions of a honest block branch (see Section 3). Furthermore,
under the two-block leading competitive criterion, this paper provides a
mathematical representation of the general tree with multiple block branches,
and shows that the main chain in the general tree can be determined easily by
means of the longest chain principle. Note that here our general tree is
different from that of the GHOST protocol given in Sompolinsky and Zohar
\cite{Som:2013, Som:2015}, since our general tree is directly built in the
competing process of multiple mining pools, and the longest chain (main
chain) can be easily determined by using the mathematical representation of
the general tree with multiple block branches.

So far, only a few studies have been done on the PoW blockchain system with multiple
mining pools. Important examples include the two different classes: (a)
Simulation, and (b) extending the Markov chain method of Eyal and Sirer
\cite{Eya:2014}. For (a) simulation, since the blockchain system with multiple
mining pools is very complicated, the simulation method becomes effective and
feasible. Leelavimolsilp et al. \cite{Leel:2018} used simulation to provide a
preliminary investigation on the selfish mining strategy adopted by multiple
miners, and analyzed the relative reward, the power threshold of selfish
miners, and the safety level of the Bitcoin system. Under the assumptions of
\cite{Leel:2018}, Leelavimolsilp et al. \cite{Leel:2019} further studied the
effectiveness of the selfish mining strategy. For (b) extending the Markov
chain method of Eyal and Sirer \cite{Eya:2014}, readers are referred to, for
example, Liu et al. \cite{LiuH:2018}, Marmolejo-Coss\'{\i}o et al.
\cite{Marm:2019}, Bai et al. \cite{Bai:2019}, Chang \cite{Chang:2019} and Xia
et al. \cite{Xia:2021}. Note that Li et al. \cite{Li:2021} indicated that the
Markov chain given in Eyal and Sirer \cite{Eya:2014} has some deficiencies
and defects compared to the theory of Markov processes.

The GHOST protocol was first introduced by Sompolinsky and Zohar
\cite{Som:2013, Som:2015} in order to improve the security and throughput of
the Bitcoin system by using the heaviest chain principle in a tree. On the other hand,
Ethereum implements a simplified GHOST protocol which refers to orphan blocks
when observing which chain is the longest. In this case, the referenced blocks
are called the uncle blocks while the referencing blocks are called the nephew
blocks. For the uncle and nephew blocks, the Ethereum system with multiple
mining pools will be faced with two basic challenges. The first challenge is
how to set up a tree structure with multiple block branches and forked
positions, which expresses the competition process of multiple mining pools.
The second challenge is how to design the rewards of uncle blocks and nephew
blocks, which are used to increase the mining enthusiasm of the multiple
mining pools, and especially when those mining pools cannot access the
main chain. For the two challenging problems, so far there has not been a
clear answer or a better research yet. This motivates us in this paper to
explore both setting up the tree structure and designing the rewards of the
uncle and nephew blocks. By finding the mathematical representation of a
general Ethereum tree, this paper applies the law of large numbers and the
reward renewal theorem to make some key and important progress in the study of
Ethereum systems with multiple mining pools.

There have been a few works on the uncle and nephew blocks and their reward
design in Ethereum up to now. Zhang et al. \cite{ZhangZKa:2020, ZhangZKb:2020}
analyzed the benefits of selfish mining in Ethereum, and chose the maximum,
7/8 units as the uncle block reward, and 1/32 units as the nephew block
reward. Lerner \cite{Lerner:2016} found that the uncle block strategy may
cause the deliberate increase in the supply of Ethereum, thus it indirectly
reduces the value of Ethereum. Ritz and Zugenmaier \cite{Ritz:2018} set up a
Monte Carlo simulation platform to quantify how the uncle blocks affect the
probability of selfish mining. Chang et al. \cite{ChangP:2019} introduced the
uncle block attacks to discuss the incentive compatibility among the different
attacks. Werner et al. \cite{Werner:2019} formally reconstructed a Sybil
attack to exploit the uncle block distribution policy in a queue-based mining
pool. Chang et al. \cite{ChangP:2019} and Werner et al. \cite{Werner:2019}
provided the simulation analysis for the uncle blocks in Ethereum. Zhang
\cite{Zhang:2020} developed a Markov decision process model to analyze the
profitability and threshold of the three-player attacks. In addition, the
uncle block mechanism can improve the security of Ethereum systems with
multiple mining pools. For the details of the two (honest and dishonest) mining pools,
interested readers are referred to Feng and Niu \cite{Feng:2019}, Grunspan and
P\'{e}rez-Marco \cite{Grun:2020}, Kang et al. \cite{Kang:2021},
\cite{Liu:2020} and Wang et al. \cite{WangL:2019}. Comparing with the studies
above, this paper proposes a new method with two consecutive rounds of mining
competition for analyzing the uncle and nephew blocks, in which we set up a
basic relation among the uncle and nephew blocks (see Figure 11) so that the
rewards of the uncle and nephew blocks can be estimated easily (see Sections 4
and 8). Obviously, one key finding of this paper is to reveal that the uncle
blocks and the nephew blocks must appear in two different rounds of mining
competition. This is crucial and interesting in the research of Ethereum.

Based on the above analysis, we summarize the main contributions of this paper
as follows:

\begin{itemize}
\item[1.] Under the two-block leading competitive criterion, we describe an
Ethereum system with one honest mining pool and multiple dishonest mining
pools, set up a general tree with multiple block branches where the multiple
dishonest block branches can fork at different positions of the one honest
block branch, and provide a mathematical representation of the general tree.
By using the mathematical representation of tree, we provide an effective
method to easily determine the main chain by means of the longest chain
principle. (See Sections 2 and 3)

\item[2.] By using the mathematical representation of tree, we propose a
two-stage mechanism to find the uncle and nephew blocks and then design the
uncle block and nephew block rewards in two consecutive rounds of mining
competition. (See Sections 4 and 8)

\item[3.] We apply the law of large numbers to study some key probabilities in
the Ethereum system with multiple mining pools, and define and compute some
key ratios: The main chain ratio, the orphan block ratio, the uncle block
ratio, the stale block ratio, and the chain quality. Note that the key ratios
are necessary and useful in the security analysis of Ethereum systems with
multiple mining pools. (See Sections 5 and 6)

\item[4.] On the one hand, we provide expression for the long-term growth rate
of blockchain by using the renewal reward theory, which is one of the most
important indicators for the Ethereum system, where the growth rate of
blockchain is the block number in all the main chains increasing per unit
time. (See Section 7). On the other hand, once the uncle and nephew block
rewards are determined, we provide expressions both for the long-term reward
allocation and for the long-term reward allocation rate to each mining pool by
using the renewal reward theory. (See Sections 8 and 9)

\item[5.] We use some simulation experiments to discuss the Ethereum system
with one honest mining pool and two dishonest mining pools, verify how the key
probabilities of Ethereum are obtained approximately by using the law of large
numbers, and analyze the performance measures of the Ethereum system, for
example, the long-term growth rate of blockchain, the long-term reward
allocation and the long-term reward allocation rate to each mining pool. We
show that the approximative computation of the performance measures of the
Ethereum system can have a faster convergence. (See Section 10)
\end{itemize}

The remainder of this paper is organized as follows. Section \ref{Sec-2:model}
describes an Ethereum system with multiple mining pools and provides the
mathematical representation of a general tree with multiple block branches.
Section \ref{Sec-3:example} gives some examples with one honest mining pool
and two dishonest mining pools for analyzing the mathematical representation
of the general tree. Section \ref{Sec-4:uncle} introduces a classification of
blocks, gives some conditions under which the orphan block can become an uncle
block, and provides a two-stage mechanism to determine the uncle and nephew
blocks. Section \ref{Sec-5:largenumber} studies some key probabilities of
Ethereum by using the law of large numbers. Section \ref{Sec-6:ratios} defines
some key ratios for the general tree with multiple block branches. Section
\ref{Sec-7:growthrate} applies the renewal reward theory to discuss the
long-term growth rate of blockchain. Section \ref{Sec-8:reward} provides a
long-term reward allocation to each mining pool by means of the renewal reward
theory. Section \ref{Sec-9:rewardrate} applies the renewal reward theory to
study the reward allocation rates among the multiple mining pools. Section
\ref{Sec-10:simulation} conducts simulation experiments to analyze the
performance measures of the Ethereum system. Section \ref{Sec-11:conclusion}
provides some concluding remarks.

\section{Model Description}

\label{Sec-2:model}

In this section, we describe an Ethereum system with multiple mining pools.
The mining competition among the multiple mining pools directly leads to a
general tree forked at the different positions of the honest block branch by
the multiple dishonest block branches. Based on this, one of our key findings
is to provide a mathematical representation for such a tree with multiple
block branches. In addition, we introduce some mathematical notations used in
our later study.

In a PoW Ethereum with multiple mining pools, the mining competition among the
multiple mining pools is a main way to build the blockchain by means of
solving the PoW mathematical puzzles. In the process of mining competition,
some transactions are first packaged as a block with finite sizes (see Li et
al. \cite{Li:2018}), and then the block needs a nonce which is found through
solving the PoW mathematical puzzle among the multiple mining pools, in which
the success of one mining pool is based on its mining power in proportion to the
total mining power. Once the nonce is solved by one mining pool and is written
into the block, then the block is successfully mined so that it can be pegged
to the block branch of this mining pool.

\begin{figure}[ptbh]
\centering                  \includegraphics[width=12.5cm]{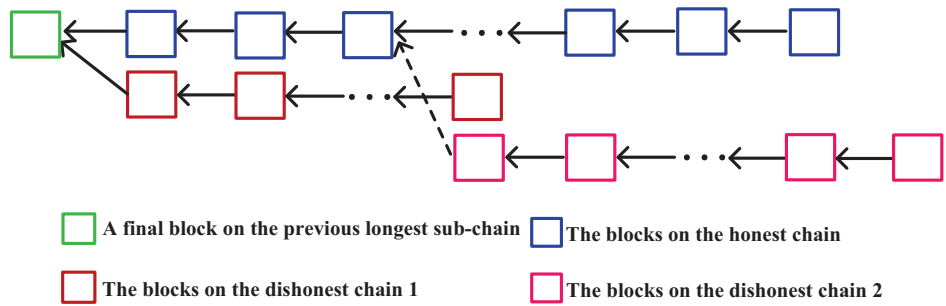}
\caption{A tree with multiple sub-chains}%
\label{Fig-1}%
\end{figure}

In the multiple mining pools, when the last round of mining competition ends,
it is easy to see the honest mining pool is major, so that it can begin to
mine from the final block on the previous sub-chain. However, the first
block mined by a dishonest mining pool can be connected to any position of the
honest sub-chain, including the final block in the last round of mining
competition, while it cannot be connected to one dishonest sub-chain mined by
the other dishonest mining pools. Based on this, we describe the tree
structure with multiple sub-chains. See Figure \ref{Fig-1} for more details.
Clearly, it is the key to provide a mathematical representation for the tree
given in Figure \ref{Fig-1}.

Note that the two-block leading competitive criterion was first proposed in Li
et al. \cite{Li:2021} in order to discuss the blockchain selfish mining with
two mining pools. In this paper, such a two-block leading competitive
criterion is further extended to the selfish mining with multiple mining
pools. Based on this, we can set up the termination rule of mining competition
among the multiple mining pools.

Now, we provide model description for the PoW Ethereum system with multiple
mining pools as follows:

\textbf{(1) Structure of PoW Ethereum system: }There are $m+1$ mining pools in
the PoW Ethereum system, where there are one honest mining pool and $m$
dishonest mining pools for $m\geq1$. For simplicity, we regard
the blocks or sub-chain mined by the honest mining pool as the honest
blocks or sub-chain, and the blocks or sub-chains mined by dishonest mining
pools as the dishonest blocks or sub-chains.

\textbf{(2) Honest mining pool: }The honest mining pool in Ethereum follows
the PoW protocol with the two-block leading competitive criterion (also see
\textbf{4-a} below). Once a block is mined by the honest mining pool, the
complete information of this block is immediately broadcasted to the entire
P2P network so that each dishonest mining pool can monitor its block
information. Thus, all the dishonest mining pools can learn about the
length of the honest sub-chain in a timely manner.

\textbf{(3) Dishonest mining pools: }The dishonest mining pools in Ethereum
follow the PoW protocol with the two-block leading competitive criterion (also
see \textbf{4-b} below), and they can carry out various selfish mining
attacks. That is, when the dishonest mining pools launch selfish attack, the
blocks mined by each dishonest mining pool may not be immediately broadcasted
to the entire P2P network. In this case,only a part of the block branch
mined by one dishonest mining pool may be pegged on the blockchain when this
dishonest mining pool can set up the main chain; while another part of the
block branch is left to keep the mining competitive advantage of this
dishonest mining pool in the next round of mining competition. Thus, the
honest mining pool and the other dishonest mining pools cannot know the
accurate information of blocks mined by this dishonest mining pool.

\textbf{(4) The two-block leading competitive criterion and its modification: }

\textbf{(4-a)} If the honest mining pool takes the lead in mining and as long
as the number of blocks mined by the honest mining pool is 2 blocks ahead of
the second-longest sub-chain mined by one of the dishonest mining pools, then
the round of mining competition terminates immediately, and the longest
sub-chain mined by the honest mining pool becomes the main chain which is
pegged onto the blockchain, while all the other sub-chains mined by the
dishonest mining pools become the chains of orphan blocks, all of which are
returned to the transaction pool without any new transaction fee.

\textbf{(4-b)} If one dishonest mining pool takes the lead in mining and as
long as the number of blocks mined by this dishonest pool is at least 2 blocks
ahead of the second-longest sub-chain among the other mining pools, then the
sub-chain mined by this dishonest mining pool becomes the main chain, while
this dishonest pool may release only a part of the main chain into the
blockchain under a basic condition that the part of the main chain is still at
least 2 blocks ahead of the second-longest sub-chain, and another part of the
main chain is reserved for the next round of mining competition in order to
keep the mining competitive advantage of this dishonest mining pool. Once the
part of the main chain begins to peg onto the blockchain, the round of mining
competition terminates immediately, and all the other sub-chains become the
chains of orphan blocks, all of which are returned to the transaction pool
without any new transaction fee.

\textbf{(5) A key mathematical representation for the }$m+1$\textbf{
sub-chains of the tree: }

\begin{figure}[ptbh]
\centering                  \includegraphics[width=12.5cm]{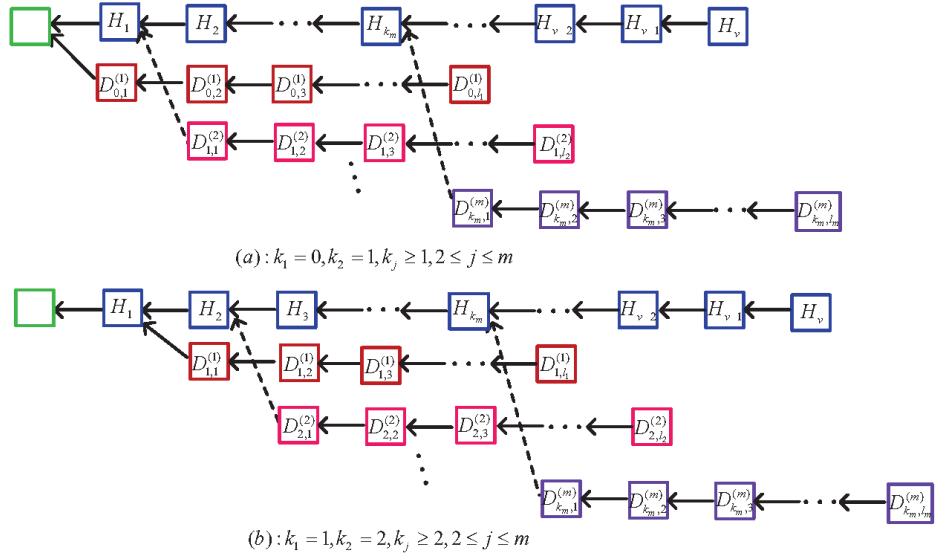}
\caption{A mathematical representation for the tree with $m+1$ sub-chains}%
\label{Fig-2}%
\end{figure}

Let $L=\left\{  L_{0},L_{1},L_{2},\ldots,L_{m}\right\}  $ denote the tree with
$m+1$ sub-chains mined by one honest mining pool and\ $m$ dishonest mining
pools, where $L_{0}$ is the sub-chain mined by the honest mining pool and
$L_{i}$ is the sub-chain mined by the $i$th dishonest mining pool for
$i=1,2,\ldots,m$. We write%
\[
L_{0}=\left\{  H_{1},H_{2},\ldots,H_{v}\right\}  ,
\]
where $H_{l}$ denotes the $l$th block mined by the honest mining pool for
$l=1,2,\ldots,v$. That is, there are $v$ blocks in the sub-chain mined by the
honest mining pool. Similarly, we write
\[
L_{i}=\left\{  H_{1},H_{2},\ldots,H_{k_{i}};D_{k_{i},1}^{(i)},D_{k_{i}%
,2}^{(i)},\ldots,D_{k_{i},l_{i}}^{(i)}\right\}  ,
\]
where $k_{i}$ represents the number of honest blocks which have been mined by
the honest mining pool before the $i$th dishonest mining pool begins to fork
after the block $H_{k_{i}},$ such that a new sub-chain with $l_{i}$ blocks is
mined by the $i$th dishonest mining pool, where $D_{k_{i},l}^{(i)}$ denotes
the $l$th block mined by the $i$th dishonest mining pool for $l=1,2,\ldots
,l_{i}$,. Figure \ref{Fig-2} provides a more intuitive understanding for the
$m+1$ sub-chains of the tree.

\textbf{(6) A mathematical expression for the mining terminative rules: }

From $L_{0}=\left\{  H_{1},H_{2},\ldots,H_{v}\right\}  $ and $L_{i}=\left\{
H_{1},H_{2},\ldots,H_{k_{i}};D_{k_{i},1}^{(i)},D_{k_{i},2}^{(i)}%
,\ldots,D_{k_{i},l_{i}}^{(i)}\right\}  $ for $i=1,2,\ldots,m$, we write
\[
\Omega=\left\{  v\right\}  \cup\left\{  \left(  k_{1},l_{1}\right)  ,\left(
k_{2},l_{2}\right)  ,\ldots,\left(  k_{m},l_{m}\right)  \right\}  .
\]
Further, we write%
\[
S=\left\{  v,k_{1}+l_{1},k_{2}+l_{2},\ldots,k_{m}+l_{m}\right\}  ,
\]
where, $v$ is the length of sub-chain mined by the honest mining pool, and
$k_{i}+l_{i}$ is the generalized sub-chain mined by the $i$th dishonest mining
pool for $i=1,2,\ldots,m$. Based on this, the elements of the set $S$ are
sorted from the largest to the smallest as follows:
\[
\omega_{1}\geq\omega_{2}\geq\cdots\geq\omega_{m}\geq\omega_{m+1}.
\]

\textbf{(6-a)} If $\omega_{1}$ comes from the honest mining pool and
$\omega_{1}-\omega_{2}=2$, then this round of mining competition terminates
immediately; the sub-chain mined by the honest mining pool is the main chain
and is pegged onto the blockchain, while all the sub-chains mined by the $m$
dishonest mining pools are returned to the transaction pool.

\textbf{(6-b)} If $\omega_{1}$ comes from the $i$th dishonest mining pool and
$\omega_{1}-\omega_{2}\geq2$, then this round of mining competition may
terminate, the sub-chain mined by the $i$th dishonest mining pool is the main
chain, and the part with $\varphi_{i}$ blocks for $\omega_{2}+2\leq\varphi
_{i}\leq\omega_{1}$ of the main chain mined by the $i$th dishonest mining pool
begins to peg onto the blockchain, while another part with $\omega_{1}%
-\varphi_{i}$ blocks of the main chain is reserved for the next round of
mining competition in order to keep the mining competitive advantage of the
$i$th dishonest mining pool. In this case, all the sub-chains mined by the
other (honest and dishonest) mining pools are returned to the transaction pool.

\textbf{(6-c)} If $0\leq\omega_{1}-\omega_{2}\leq1$, then this round of mining
competition cannot terminate and the mining pools continue to mine until the
two-block leading competitive criterion is satisfied.

\textbf{(7) The mining rewards: }When a round of mining competition
terminates, the main chain is pegged onto the blockchain, and all the other
sub-chains become orphan blocks which are returned to the transaction pool,
waiting for the next round of mining competition.

\textbf{(7-a)} If the main chain comes from the honest mining pool and
$L_{0}=\left\{  H_{1},H_{2},\ldots,H_{v}\right\}  $, then the honest mining
pool obtains the rewards of $v$ blocks.

\textbf{(7-b)} If the main chain comes from the $i$th dishonest mining pool,
and
\[
L_{i}=\left\{  H_{1},H_{2},\ldots,H_{k_{i}};D_{k_{i},1}^{(i)},D_{k_{i}%
,2}^{(i)},\ldots,D_{k_{i},l_{i}}^{(i)}\right\}  ,
\]
then the honest mining pool obtains the rewards of $k_{i}$ blocks. Let
$\omega_{1}=k_{i}+l_{i}$, then the $i$th dishonest mining pool obtains the
rewards of $\varphi_{i}$ blocks for $\omega_{2}+2\leq\varphi_{i}\leq\omega
_{1}$. Note that the $\omega_{1}-\varphi_{i}$ blocks of the main chain cannot
be pegged onto the blockchain, thus they cannot lead to any reward for the $i$th
dishonest mining pool.

In addition, the uncle block rewards and associated reference rewards will be
assumed and discussed in Section \ref{Sec-4:uncle}.

\section{Examples for the Tree Representation}

\label{Sec-3:example}

This section provides some examples to analyze the mathematical representation
of the general tree with one honest mining pool and two dishonest mining
pools. Note that the mathematical representation of tree plays a key role in
the study of PoW Ethereum systems with multiple mining pools.

In the PoW Ethereum system with multiple mining pools, the tree with multiple
sub-chains expresses the mining competition process among the honest mining
pool and the $m$ dishonest mining pools. The following theorem provides an
essential feature of the tree with multiple sub-chains.

\begin{The}
In a round of mining competition and from the two-block leading competitive
criterion, we have

(a) if the sub-chain of the $i$th dishonest mining pool is the first forked
sub-chain among all the $m$ dishonest mining pools, then either $k_{i}=0$ or
$k_{i}=1$; and

(b) if the sub-chain of the $i$th dishonest mining pool is not the first
forked sub-chain among all the $m$ dishonest mining pools, then $k_{i}\geq1$.
\end{The}

\textbf{Proof.} (a) We provide the proof by contradiction. We assume that the
sub-chain of the $i$th dishonest mining pool is the first forked sub-chain
among all the $m$ dishonest mining pools, and $k_{i}\geq2$. In this case, by
using the two-block leading competitive criterion, it is easy to see that the
honest mining pool is at least 2 blocks ahead of the second-longest sub-chain
among the other mining pools, the honest mining pool immediately terminates
this round of mining competition. Obviously, it is impossible that the $i$th
dishonest mining pool can fork at $k_{i}\geq2$. Therefore, this gives that
either $k_{i}=0$ or $k_{i}=1$.

(b) If the sub-chain of the $i$th dishonest mining pool is not the first
forked sub-chain among all the $m$ dishonest mining pools, then there must be
a $j$th dishonest mining pool who is the first one to fork at the tree for
$j\neq i$. By using (a), either $k_{j}=0$ or $k_{j}=1$. In this case, our
discussion has two different cases as follows:

(i) When $k_{j}=0$, since the sub-chain of the $i$th dishonest mining pool is
not the first forked sub-chain among all the $m$ dishonest mining pools, it is
easy to see that the $i$th dishonest mining pool can fork at a later position
than the dishonest pool $j$, this gives $k_{i}>k_{j}=0$, i.e., $k_{i}\geq1$.
In this case, $k_{i}\geq1$.

(ii) When $k_{j}=1$, since the sub-chain of the $i$th dishonest mining pool is
not the first forked sub-chain among all the $m$ dishonest mining pools, it is
easy to see that the $i$th dishonest mining pool can fork at a later position
than the dishonest pool $j$, this gives $k_{i}>k_{j}=1$, i.e., $k_{i}\geq2$.
In this case, $k_{i}\geq2$.

From the above two cases, we get that $k_{i}\geq1.$ This completes the proof.
$\square$

In the remainder of this section, we analyze some examples of the honest
mining pool and one dishonest mining pool setting up the main chain, respectively.

\subsection{The honest mining pool sets up the main chain}

If the honest mining pool sets up the main chain in a round of mining
competition, this subsection provides four different examples to express the
tree with at most three sub-chains.

\begin{figure}[h]
\centering         \includegraphics[width=3.5cm]{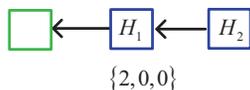}  \caption{The
first case}%
\label{Fig-3}%
\end{figure}

\textbf{Tree one}: The $1$st and $2$nd dishonest mining pools have not mined
any block yet, while the honest mining pool has mined two blocks in a round of
mining competition. Thus the round of mining competition is terminated
immediately. See Figure \ref{Fig-3}.

\begin{figure}[h]
\centering          \includegraphics[width=9.5cm]{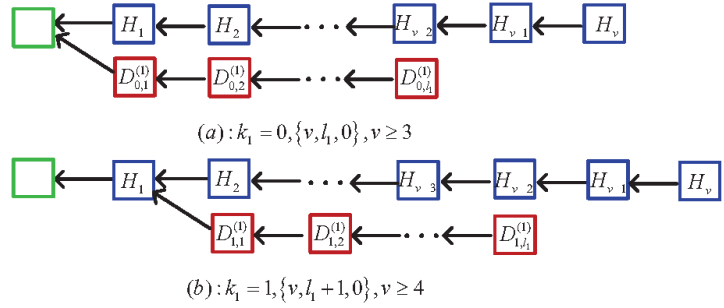}  \caption{The
second case}%
\label{Fig-4}%
\end{figure}

\textbf{Tree two}: The sub-chain lengths of the $1$st and $2$nd dishonest
mining pools and the honest mining pool are $l_{1}$, $0$ and $v$,
respectively. If $k_{1}=0$, then the round of mining competition ends due to
$l_{1}=v-2$. If $k_{1}=1$, then the round of mining competition ends due to
$l_{1}+1=v-2$. See Figure \ref{Fig-4}.

\begin{figure}[h]
\centering      \includegraphics[width=10cm]{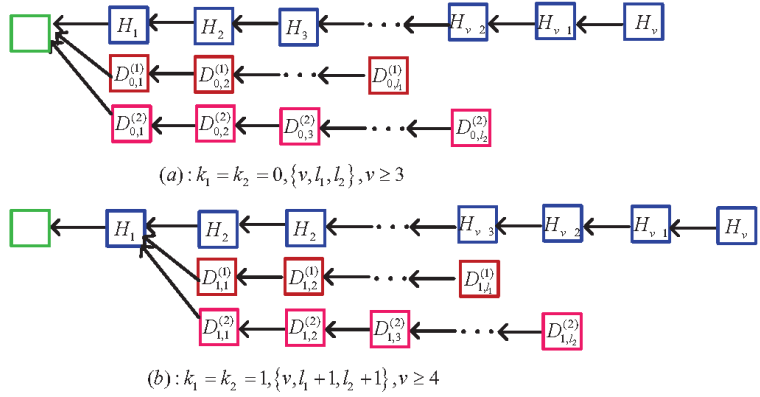}  \caption{The third
case}%
\label{Fig-5}%
\end{figure}

\textbf{Tree three}: The sub-chain lengths of the two dishonest mining pools
are not $0$, and the two sub-chains fork at the same position. If $k_{1}%
=k_{2}=0$, the round of mining competition ends at the condition under which
either $l_{1}=v-2,1\leq$ $l_{2}\leq v-2$ or $l_{2}=v-2,1\leq$ $l_{1}\leq v-2$.
If $k_{1}=k_{2}=1$, the round of mining competition ends at the condition
under which either $l_{1}+1=v-2,2\leq$ $l_{2}+1\leq v-2$ or $l_{2}%
+1=v-2,2\leq$ $l_{1}\leq v-2$. See Figure \ref{Fig-5}.

\begin{figure}[h]
\centering            \includegraphics[width=11cm]{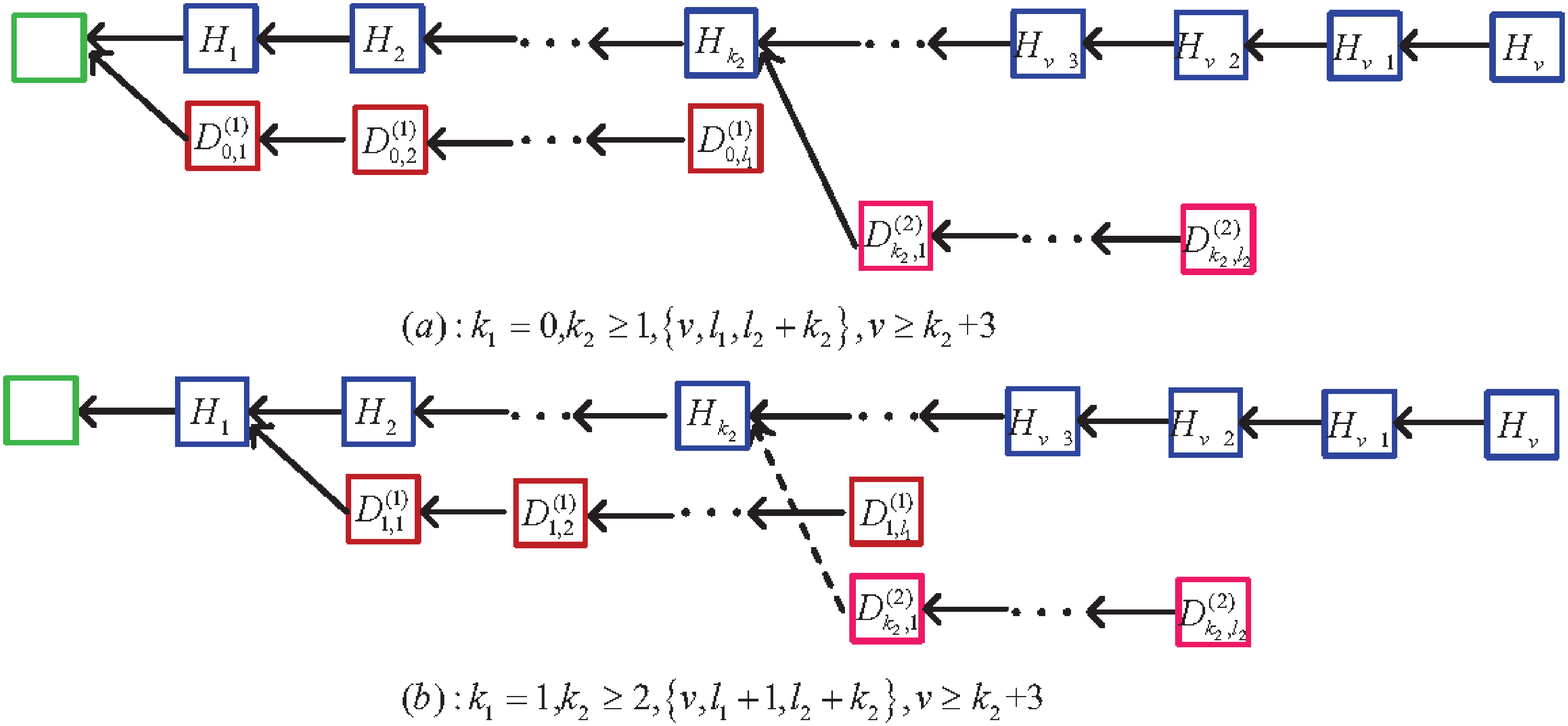}  \caption{The
fourth case}%
\label{Fig-6}%
\end{figure}

\textbf{Tree four}: The sub-chain lengths of the two dishonest mining pools
are not 0, and the two sub-chains fork at two different positions. If
$k_{1}=0,k_{2}\geq1$, then the round of mining competition ends at the
condition under which either $l_{1}=v-2,2\leq$ $l_{2}+k_{2}\leq v-2$ or
$l_{2}+k_{2}=v-2,1\leq$ $l_{1}\leq v-2$. If $k_{1}=1,k_{2}\geq2$, then the
round of mining competition ends at the condition under which either
$l_{1}+1=v-2,3\leq$ $l_{2}+k_{2}\leq v-2$ or $l_{2}+k_{2}=v-2,2\leq$
$l_{1}+1\leq v-2$. See Figure \ref{Fig-6}.

\subsection{One dishonest mining pool sets up the main chain}

If one dishonest mining pool sets up the main chain in a round of mining
competition, this subsection provides four different examples to express the
tree with at most three sub-chains.

\begin{figure}[h]
\centering         \includegraphics[width=7.5cm]{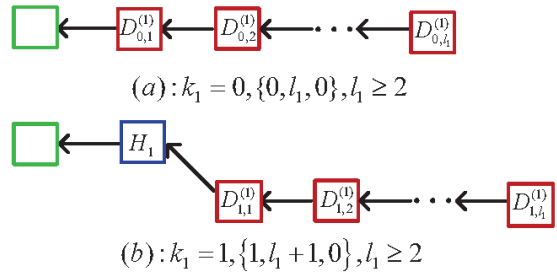}  \caption{The
first case}%
\label{Fig-7}%
\end{figure}

\textbf{Tree one}: The sub-chain length of the $1$st dishonest mining pool is
at least $2$, but the sub-chain lengths of the $2$nd dishonest mining pool and
the honest mining pool after the $1$st dishonest mining pool forks are $0$.
See Figure \ref{Fig-7}.

\begin{figure}[h]
\centering         \includegraphics[width=8.5cm]{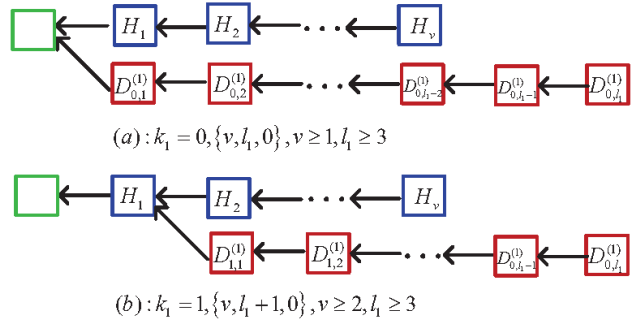}  \caption{The
second case}%
\label{Fig-8}%
\end{figure}

\textbf{Tree two}: The sub-chain lengths of the $1$st dishonest mining pool
and the honest mining pool after the $1$st dishonest mining pool forks are
positive, but the sub-chain length of the $2$nd dishonest mining pool is $0$.
If $k_{1}=0$, then the round of mining competition ends at the condition under
which $l_{1}\geq v+2,v\geq1$. If $k_{1}=1$, then the round of mining
competition ends at the condition under which $l_{1}+1\geq v+2,v\geq2$. See
Figure \ref{Fig-8}.

\begin{figure}[h]
\centering            \includegraphics[width=9.5cm]{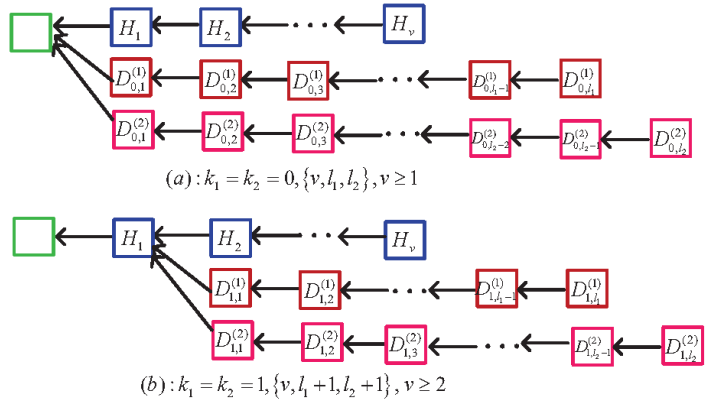}  \caption{The
third case}%
\label{Fig-9}%
\end{figure}

\textbf{Tree three}: The $1$st and $2$nd dishonest mining pools fork at the
same position, and $l_{1}\neq0,$ $l_{2}\neq0$; while the honest mining pool
has a positive sub-chain length after the $1$st and $2$nd dishonest mining
pools fork, thus, $v-k_{1}\geq1$. If $k_{1}=k_{2}=0$, then the round of mining
competition ends with the condition under which either $l_{1}\geq\max\left\{
v+2,l_{2}+2\right\}  $ or $l_{2}\geq\max\left\{  v+2,l_{1}+2\right\}  $. If
$k_{1}=k_{2}=1$, then the round of mining competition ends at the condition
under $l_{1}+1\geq\max\left\{  v+2,l_{2}+3\right\}  $ or $l_{2}+1\geq
\max\left\{  v+2,l_{1}+3\right\}  $. See Figure \ref{Fig-9}.

\begin{figure}[h]
\centering                       \includegraphics[width=11cm]{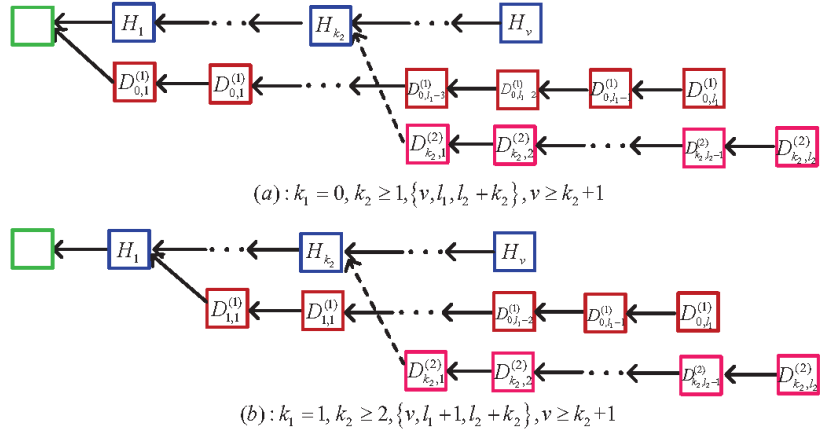}
\caption{The fourth case}%
\label{Fig-10}%
\end{figure}

\textbf{Tree four}: The two dishonest mining pools do not fork at the same
position, and $l_{1}\neq0,$ $l_{2}\neq0$; while the honest mining pool has a
positive sub-chain length after the $1$st dishonest mining pool forks, i.e.,
$v-k_{1}\geq1$. Now, we consider two cases: $k_{1}=0$, $k_{2}\geq1$ and
$k_{1}=1$, $k_{2}\geq2$. If the $1$st dishonest mining pool sets up the main
chain, then the round of mining competition ends at the condition under which
either%
\[
l_{1}\geq\max\left\{  v+2,l_{2}+k_{2}+2\right\}  ,k_{1}=0,k_{2}\geq1;
\]
or%
\[
l_{1}+1\geq\max\left\{  v+2,l_{2}+k_{2}+2\right\}  ,k_{1}=1,k_{2}\geq2.
\]
If the $2$nd dishonest mining pool sets up the main chain, then the round of
mining competition ends at the condition under which either%
\[
l_{2}+k_{2}\geq\max\left\{  v+2,l_{1}+2\right\}  ,k_{1}=0,k_{2}\geq1;
\]
or%
\[
l_{2}+k_{2}\geq\max\left\{  v+2,l_{1}+3\right\}  ,k_{1}=1,k_{2}\geq2.
\]
See Figure \ref{Fig-10}.

\section{The Uncle Blocks and Reward Design}

\label{Sec-4:uncle}

In this section, we introduce a classification of blocks, and give some
conditions under which the orphan block can become an uncle block.
Furthermore, we provide a two-stage mechanism to determine the uncle blocks
and the nephew blocks.

\begin{figure}[h]
\centering             \includegraphics[width=10.5cm]{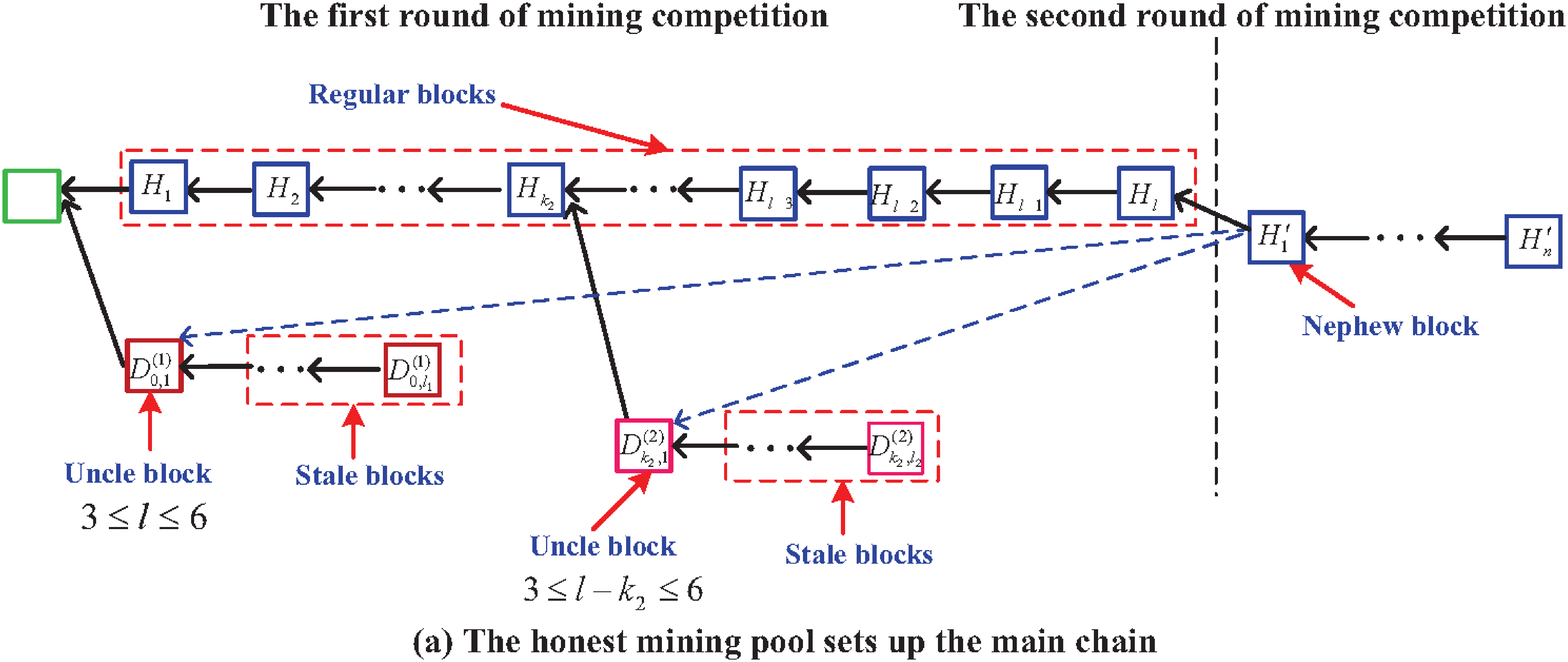}
\includegraphics[width=10.5cm]{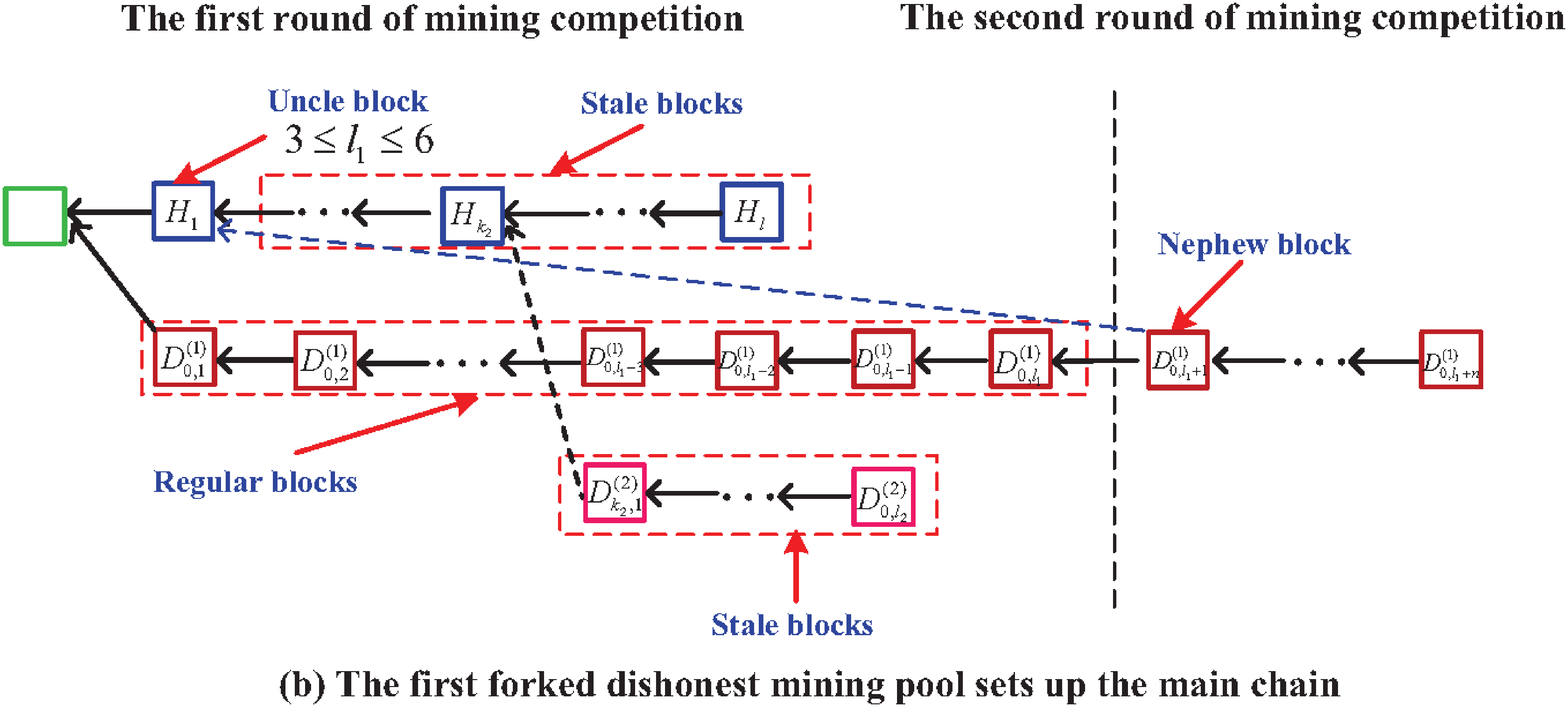}
\includegraphics[width=10.5cm]{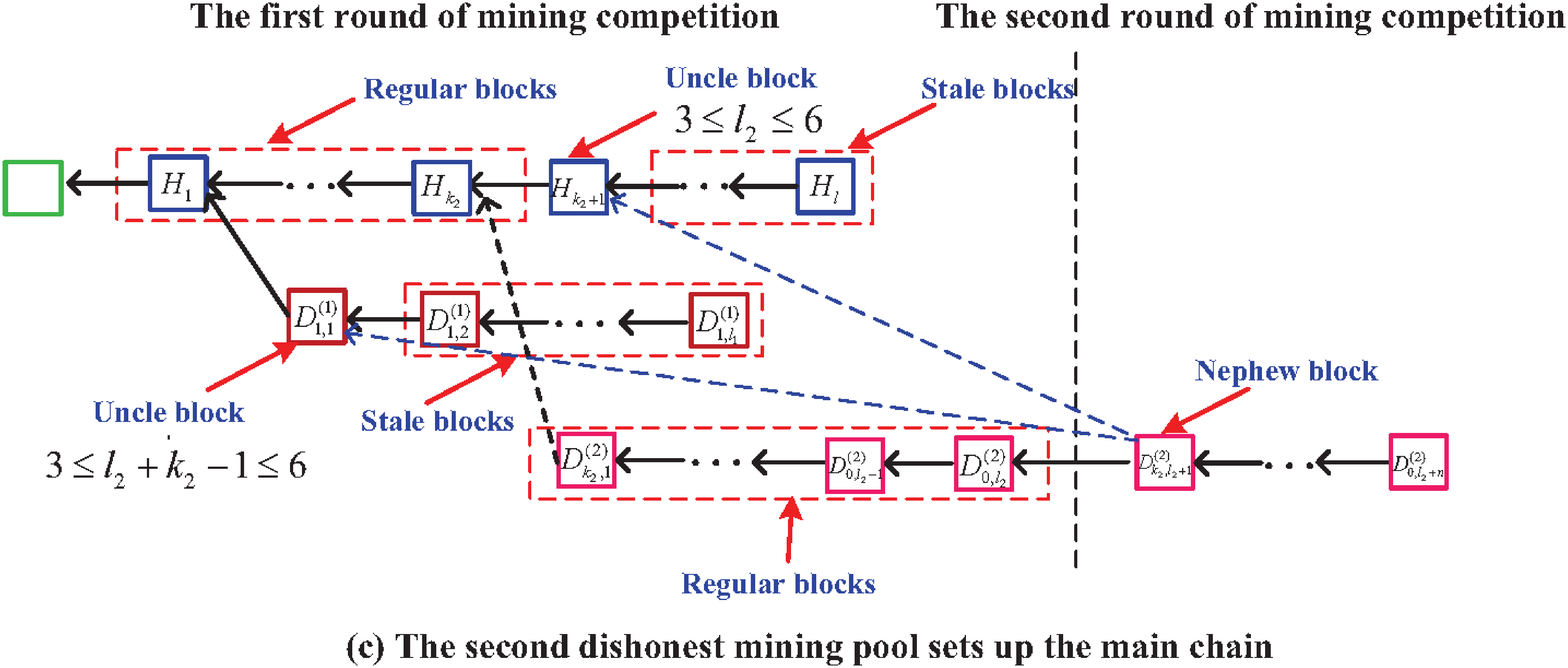}  \caption{A classification of
blocks}%
\label{Fig-11}%
\end{figure}

In the multiple sub-chains of the tree corresponding to the PoW Ethereum
systems with multiple mining pools, we divide the blocks into five different
types: Regular blocks, orphan blocks, uncle blocks, stale blocks, and nephew
blocks. Also, the orphan blocks are further divided into the uncle blocks and
the stale blocks, if any. See Figure \ref{Fig-11} for an intuitive understanding.

The regular block is a block of the main chain. The uncle block must satisfy a
key condition under which the distance between the uncle block and the nephew
block does not exceed $7$ blocks. The nephew block is the first block in the
next round of mining competition, and it is used to determine the uncle blocks
from the sub-chains of the tree, as seen in Figure \ref{Fig-11}. The stale
blocks follow from an uncle block in a sub-chain of the tree, or the stale
blocks are all the blocks in each sub-chain of orphan blocks.

For the mining pools in Ethereum, their reward include 4 parts: A regular
block reward is given by Ethereum, transaction costs (i.e. gas fee), an uncle
block reward, and a nephew reference reward. In this paper, we take the unit
of economic measure as ``block'', in other words, the regular reward given by
the Ethereum systems is assumed to be 1.

In the PoW Ethereum system with multiple mining pools, the main purpose of
introducing the uncle blocks is to inspire the mining enthusiasm of each
mining pool, especially for those mining pools who cannot set up the main
chain. In this case, they cannot obtain any reward but have to cover some
costs associated with the mining processes, such as electricity, equipment
investment, maintenance costs, staff salary, management fees and so on. Thus,
providing the uncle block reward and the nephew reference reward can be a
necessary and valuable support for the multiple mining pools that take an
active part in each round of mining competition.

To show how to allocate the uncle block reward and the nephew reference
reward, we provide a two-stage reward allocation mechanism through analyzing
the position among the uncle blocks and the nephew blocks. For simplicity of
analysis, the uncle blocks are determined at the moment that a round of mining
competition has been over, and a nephew block (or the first block) is just
generated in the next round of mining competition. That is, to determine the
uncle blocks, we first need to have the nephew block in order that we can
judge that the distance between the uncle block and the nephew block does not
exceed $7$ blocks. Note that the nephew block comes from two different cases:
(i) If all the blocks of the main chain are released at the ending moment of
the last round of mining competition, then the nephew block is the first block
mined by one of the $m+1$ mining pools in the next round of mining
competition. (ii) If there are some blocks of the main chain reserved at the
ending moment of the last round of mining competition, then these reserved
blocks must come from one dishonest mining pool, and the nephew block is the
first one of the reserved blocks.

It is worthwhile to note that such a reward allocation is well related to the
distance between the uncle block and the nephew block. Table 1 shows the uncle
block rewards at some different distances.

\begin{table}[h]
\caption{Table of uncle rewards at different distances }%
\label{tab:tab1}%
\center   { {\small \setlength{\abovecaptionskip}{0cm}%

\setlength{\belowcaptionskip}{-0.2cm}  {
\begin{tabular}
[c]{ccccccc}%
\toprule[1pt]  \textbf{Distance} & 1 & 2 & 3 & 4 & 5 & 6\\
\specialrule{1pt}%
{1pt}{1pt}  \textbf{Uncle reward}%

& 7/8 & 6/8 & 5/8 & 4/8 &
3/8 & 2/8\\
\bottomrule[1pt]  &  &  &  &  &  &
\end{tabular}%

}}}
\par
{\small \vspace{-20pt} }\end{table}

In what follows we discuss some conditions under which the orphan block can
become an uncle block. Our description contains two different cases as follows:

\textbf{Case one:} If the honest mining pool sets up the main chain, then one
of the orphan blocks mined by the $i$th dishonest mining pool becomes an uncle
block and must satisfy the following two conditions (a-1) and (a-2), also see (a)
of Figure \ref{Fig-11}.

(a-1) The distance between the orphan block and its corresponding nephew block
does not exceed 7 blocks; and

(a-2) the orphan block is the first block $D_{k_{i},1}^{(i)}$ mined by the
$i$th dishonest mining pool.

\textbf{Case two:} If the $i$th dishonest mining pool sets up the main chain,
then either one of the orphan blocks mined by the $j$th dishonest mining pool
for $j\neq i$ or one of the orphan blocks mined by the honest mining pool
becomes an uncle block and must satisfy the following two conditions (b-1) and
(b-2), also see (b) and (c) of Figure \ref{Fig-11}.

(b-1) The distance between the orphan block and its corresponding nephew block
does not exceed 7 blocks; and

(b-2) either the orphan block is the first block $D_{k_{j},1}^{(j)}$ mined by
the $j$th dishonest mining pool or the orphan block is the first block
$H_{k_{i}+1}$ mined by the honest mining pool.

In addition to the uncle reward given in a round of mining competition, we
still need to provide a reward to the nephew block, called the nephew
reference reward, which is given in the next round of mining competition. A
nephew block can obtain the nephew reference reward: $N_{\text{U}}/32$, where
$N_{\text{U}}$ is the number of uncle blocks in this round of mining
competition. Note that, our above reward allocation is designed in two
consecutive rounds of mining competition, and, thus, is called a two-stage reward
allocation mechanism.

\section{The Law of Large Numbers}

\label{Sec-5:largenumber}

In this section, we apply the law of large numbers to study some key
probabilities, which are necessary and useful in our later study, such as
the key ratios of Ethereum, the growth rate of blockchain, the reward
allocation rates among the mining pools, and so forth.

In a round of mining competition, both the honest mining pool and one of the
dishonest mining pools are possible to set up the main chain. Now, we provide
an analysis for the number of main chains that can be set up by either the
honest mining pool or the $\zeta$th dishonest mining pool. During the $N$
rounds of mining competition, we denote by $N_{\text{H}}$ and $N_{\zeta}$ the
numbers that the honest mining pool and the $\zeta$th dishonest mining pool
can set up the main chain, respectively. It is clear that $0\leq N_{\text{H}%
}\leq N$, $0\leq N_{\zeta}\leq N$, and $N_{\text{H}}+\sum_{\zeta=1}%
^{m}N_{\zeta}=N$. Thus we have%
\[
\frac{N_{\text{H}}}{N}+\sum_{\zeta=1}^{m}\frac{N_{\zeta}}{N}=1\text{.}%
\]

In the PoW Ethereum system with multiple mining pools, the competitively
mining processes can be repeated round after round, as we repeat by the
similar experiments round after round under the same conditions. Thus, we can
apply the law of large numbers to study the frequencies: $N_{\text{H}}/N$ and
$N_{\zeta}/N$, and to show that they can steadily approach some fixed values
for $\zeta=1,2,\ldots,m$. The following two theorems are obvious by the law of
large numbers, and their proof are easy and omitted here for brevity.

\begin{The}
In the PoW Ethereum system with multiple mining pools, by using the law of
large numbers, as $N\rightarrow\infty$, we have%
\[
\frac{N_{\text{H}}}{N}\rightarrow\mathbf{p}_{\text{H}},\text{ \ a.s.,}%
\]
and for $\zeta=1,2,\ldots,m,$%
\[
\frac{N_{\zeta}}{N}\rightarrow\mathbf{p}_{\zeta},\text{ \ a.s..}%
\]
Also, it is clear that%
\[
\mathbf{p}_{\text{H}}+\sum_{\zeta=1}^{m}\mathbf{p}_{\zeta}=1.
\]
\end{The}

When the honest mining pool sets up the main chain, during the $N$ rounds of
mining competition, we denote by $N_{\text{H},\text{H}}^{\left(
\text{N}\right)  }$, $N_{\text{H},i}^{\left(  \text{N}\right)  }$ and
$N_{\text{H},i}^{\left(  \text{U}\right)  }$ the numbers that the honest
mining pool and the $i$th dishonest mining pool can contain one nephew block,
and the $i$th dishonest mining pool can contain one uncle block, respectively.

When the $\zeta$th dishonest mining pool sets up the main chain for
$\zeta=1,2,\ldots,m$, during the $N$ rounds of mining competition, we denote
by $N_{\zeta,\zeta}^{\left(  \text{N}\right)  }$, $N_{\zeta,\text{H}}^{\left(
\text{N}\right)  }$, $N_{\zeta,i}^{\left(  \text{N}\right)  }$, $N_{\zeta
,\text{H}}^{\left(  \text{U}\right)  }$ and $N_{\zeta,i}^{\left(
\text{U}\right)  }$ the numbers that the $\zeta$th dishonest mining pool, the
honest mining pool, and the $i$th dishonest mining pool can contain one nephew
block, and the honest mining pool and the $i$th dishonest mining pool can
contain one uncle block, respectively.

\begin{The}
In the PoW Ethereum with multiple mining pools, by using the law of large
numbers, as $N\rightarrow\infty$, for $i=1,2,\ldots,m$,%
\[
\frac{N_{\text{H},\text{H}}^{\left(  \text{N}\right)  }}{N}\rightarrow
\mathbf{q}_{\text{H},\text{H}}^{\left(  \text{N}\right)  },\text{ \ a.s.,}%
\]%
\[
\frac{N_{\text{H},i}^{\left(  \text{N}\right)  }}{N}\rightarrow\mathbf{q}%
_{\text{H},i}^{\left(  \text{N}\right)  },\text{ }\frac{N_{\text{H}%
,i}^{\left(  \text{U}\right)  }}{N}\rightarrow\mathbf{q}_{\text{H},i}^{\left(
\text{U}\right)  },\text{ \ a.s.,}%
\]
and for $i\neq\zeta$ and $i=1,2,\ldots,m$,%
\[
\frac{N_{\zeta,\zeta}^{\left(  \text{N}\right)  }}{N}\rightarrow
\mathbf{q}_{\zeta,\zeta}^{\left(  \text{N}\right)  },\text{ \ a.s.,}%
\]%
\[
\frac{N_{\zeta,\text{H}}^{\left(  \text{N}\right)  }}{N}\rightarrow
\mathbf{q}_{\zeta,\text{H}}^{\left(  \text{N}\right)  },\text{ }%
\frac{N_{\zeta,\text{H}}^{\left(  \text{U}\right)  }}{N}\rightarrow
\mathbf{q}_{\zeta,\text{H}}^{\left(  \text{U}\right)  },
\]%
\[
\frac{N_{\zeta,i}^{\left(  \text{N}\right)  }}{N}\rightarrow\mathbf{q}%
_{\zeta,i}^{\left(  \text{N}\right)  },\frac{N_{\zeta,i}^{\left(
\text{U}\right)  }}{N}\rightarrow\mathbf{q}_{\zeta,i}^{\left(  \text{U}%
\right)  },\text{ \ a.s..}%
\]
\end{The}

It is clear that $0\leq\mathbf{q}_{\text{H},\text{H}}^{\left(  \text{N}%
\right)  },\mathbf{q}_{\text{H},i}^{\left(  \text{N}\right)  }\leq1$ and
$0\leq\mathbf{q}_{\text{H},i}^{\left(  \text{U}\right)  }\leq1$ for
$i=1,2,\ldots,m;$ and $0\leq\mathbf{q}_{\zeta,\zeta}^{\left(  \text{N}\right)
},\mathbf{q}_{\zeta,\text{H}}^{\left(  \text{N}\right)  },\mathbf{q}_{\zeta
,i}^{\left(  \text{N}\right)  }\leq1$ and $0\leq\mathbf{q}_{\zeta,\text{H}%
}^{\left(  \text{U}\right)  },\mathbf{q}_{\zeta,i}^{\left(  \text{U}\right)
}\leq1$ for $i\neq\zeta$ and $i=1,2,\ldots,m$.

\section{Some Key Ratios of Ethereum}

\label{Sec-6:ratios}

This section defines some key ratios of the PoW Ethereum system with multiple
mining pools, and provides a detailed analysis for the key ratios by means of
the mathematical representation of tree with multiple sub-chains.

In a round of mining competition, the multiple mining pools use their mined
blocks to set up a tree with multiple sub-chains. From the tree, we can
classify five different types of blocks: Regular blocks, orphan blocks, uncle
blocks, nephew blocks, and stale blocks. Based on this, we can set up some key
ratios of the PoW Ethereum system with multiple mining pools. To this end, we
define some key ratios of Ethereum from two perspectives of efficiency and
benefit, such as chain quality, main chain length ratio, orphan block ratio,
uncle block ratio, and stale block ratio.

\begin{Def}
\label{Def:Ratio-1}In a round of mining competition, we define

(a) \textbf{The chain quality }$c_{\text{Q}}$\textbf{:} It is defined as the
ratio that some blocks on the main chain mined by the honest mining pool
occupy all the blocks of the main chain.

(b) \textbf{The main chain length ratio }$r_{\text{M}}$\textbf{:} It is
defined as the ratio that the number of blocks on the main chain over the
number of blocks on the tree.
\end{Def}

If the honest mining pool sets up the main chain, from $L_{0}=\left\{
H_{1},H_{2},\ldots,H_{v}\right\}  $ and $L_{i}=\left\{  H_{1},H_{2}%
,\ldots,H_{k_{i}};D_{k_{i},1}^{(i)},D_{k_{i},2}^{(i)},\ldots,D_{k_{i},l_{i}%
}^{(i)}\right\}  $ for $i=1,2,\ldots,m$, we have%
\[
c_{\text{Q,H}}=\frac{v}{v}=1
\]
and%

\[
\ r_{\text{M,H}}=\frac{v}{v+\sum\limits_{i=1}^{m}l_{i}}.
\]

If the $i$th dishonest mining pool sets up the main chain for $i=1,2,\ldots
,m$, due to the part with $\varphi_{i}$ blocks for $\omega_{2}+2\leq
\varphi_{i}\leq\omega_{1}$ of the main chain are pegged on the blockchain,
while the $\omega_{1}-\varphi_{i}$ blocks of the main chain cannot be
observed by all the other mining pools in the P2P network, then%

\[
\ c_{\text{Q,}i}=\frac{k_{i}}{k_{i}+\varphi_{i}}%
\]
and$\ $%
\[
r_{\text{M,}i}=\frac{\varphi_{i}+k_{i}}{v+\varphi_{i}+\sum\limits_{j=1,\text{
}j\neq i}^{m}l_{j}}.
\]

\begin{Def}
\label{Def:Ratio-2}In a round of mining competition, we define

(a) \textbf{The orphan block ratio }$r_{\text{O}}$\textbf{: }It is defined as
the ratio of the number of orphan blocks to the number of blocks on the tree.

(b) \textbf{The uncle block ratio }$r_{\text{U}}$\textbf{: }It is defined as
the ratio of the number of uncle blocks to the number of blocks on the tree.
\end{Def}

It is easy to see from Definitions 1 and 2 that%

\[
r_{\text{M}}+r_{\text{O}}=1
\]
and%

\[
r_{\text{O}}=r_{\text{U}}+r_{\text{S}}.
\]

In what follows, we first express the orphan block ratio $r_{\text{O}}$. To
this end, our computation needs to consider two different cases: The honest
mining pool sets up the main chain, and one dishonest mining pool sets up the
main chain.

If the honest mining pool sets up the main chain, then%

\begin{equation}
\ r_{\text{O,H}}=\frac{\sum\limits_{i=1}^{m}l_{i}}{v+\sum\limits_{i=1}%
^{m}l_{i}}. \label{O-1}%
\end{equation}
\ \

If the $i$th dishonest mining pool sets up the main chain for $i=1,2,\ldots
,m$, due to the part with $\varphi_{i}$ blocks for $\omega_{2}+2\leq
\varphi_{i}\leq\omega_{1}$ of the main chain are pegged onto the blockchain,
while the $\omega_{1}-\varphi_{i}$ blocks of the main chain cannot be
observed by all the other mining pools in the P2P network, then%

\begin{equation}
\ r_{\text{O,}i}=\frac{\left(  v-k_{i}\right)  +\sum\limits_{j=1,\text{ }j\neq
i}^{m}l_{j}}{v+\varphi_{i}+\sum\limits_{j=1,\text{ }j\neq i}^{m}l_{j}}.
\label{O-2}%
\end{equation}

Now, we compute the uncle block ratio $r_{\text{U}}$. Note that such a
computation is a little bit complicated.

To compute the uncle block ratio $r_{\text{U}}$, it is necessary to first
determine how many orphan blocks can become uncle blocks. To this end, our
computation also needs to consider two different cases: The honest mining pool
sets up the main chain, and one dishonest mining pool sets up the main chain.

\textbf{Case one: The honest mining pool sets up the main chain}

In this case, from the tree with multiple sub-chains (e.g., see (a) of Figure
\ref{Fig-11}), it is easy to determine the nephew block, and the uncle blocks
and their number. Note that $N_{\text{U,H}}$ is the number of uncle blocks in
the tree in this round of mining competition, thus the uncle block ratio is
given by%

\begin{equation}
r_{\text{U,H}}=\frac{N_{\text{U,H}}}{v+\sum\limits_{i=1}^{m}l_{i}}.
\label{U-1}%
\end{equation}

\textbf{Case two: One dishonest mining pool sets up the main chain}

In this case, we assume that the $i$th dishonest mining pool sets up the main
chain, it is clear that $\omega_{1}=k_{i}+l_{i}$. We further assume that the
part with $\varphi_{i}$ blocks for $\omega_{2}+2\leq\varphi_{i}\leq\omega_{1}$
of the main chain are pegged onto the blockchain, while another part with
$\omega_{1}-\varphi_{i}$ blocks of the main chain is left in the next round of
mining competition. It is easy to see that $D_{k_{i},\varphi_{i}+1}^{(i)}$ is
the first block in the next round of mining competition. Now, our first task
is to use the block $D_{k_{i},\varphi_{i}+1}^{(i)}$ to determine which blocks
of $H_{k_{i}+1}$ and $D_{k_{j},1}^{(j)}$ for $j\neq i$ can become uncle
blocks. If $H_{k_{i}+1}$ is an uncle block, then for $j\neq i$, $D_{k_{j}%
,1}^{(j)}$ is not an uncle block for $k_{j}\geq k_{i}+1.$

From the tree with multiple sub-chains (e.g., see (b) and (c) of Figure
\ref{Fig-11}), it is easy to determine the nephew block, and the uncle blocks
and their number. Note that $N_{\text{U,}i}$ is the number of uncle blocks in
the tree, thus the uncle block ratio is given by%

\begin{equation}
r_{\text{U,}i}=\frac{N_{\text{U,}i}}{v+\varphi_{i}+\sum\limits_{j=1,\text{
}j\neq i}^{m}l_{j}}. \label{U-2}%
\end{equation}

In the remainder of this section, we apply the law of large numbers to discuss
the key ratios of the Ethereum system.

In the $r$th round of mining competition, we denote by $c_{\text{Q}}^{(r)}$,
$r_{\text{M}}^{(r)}$, $r_{\text{O}}^{(r)}$, and $r_{\text{U}}^{(r)}$ the chain
quality, the main chain length ratio, the orphan block ratio, and the uncle
block ratio, respectively. We assume that the part with $\varphi_{i}^{\left(
r\right)  }$ blocks for $\omega_{i.2}^{\left(  r\right)  }+2\leq\varphi
_{i}^{\left(  r\right)  }\leq\omega_{i,1}^{\left(  r\right)  }$ of the main
chain mined by the $i$th dishonest mining pool are pegged onto the blockchain,
where $\omega_{i.2}^{\left(  r\right)  }$ and $\omega_{i,1}^{\left(  r\right)
}$\ are the first and second elements of the sorted set $S$ related to the
$i$th dishonest mining pool, respectively.

\begin{The}
In the PoW Ethereum system with multiple mining pools, by using the law of
large numbers, as $N\rightarrow\infty$, we have%
\[
\frac{\sum\limits_{r=1}^{N}c_{\text{Q}}^{(r)}}{N}\rightarrow\overline
{c}_{\text{Q}},\text{ \ a.s., \ \ \ \ \ }\frac{\sum\limits_{r=1}%
^{N}r_{\text{M}}^{(r)}}{N}\rightarrow\overline{r}_{\text{M}},\text{ \ a.s.,}%
\]%
\[
\frac{\sum\limits_{r=1}^{N}r_{\text{O}}^{(r)}}{N}\rightarrow\overline
{r}_{\text{O}},\text{ \ a.s., \ \ \ \ \ }\frac{\sum\limits_{r=1}%
^{N}r_{\text{U}}^{(r)}}{N}\rightarrow\overline{r}_{\text{U}},\ \text{a.s.,}%
\ \text{ }%
\]

where%
\begin{align*}
\overline{c}_{\text{Q}}  &  =\mathbf{p}_{\text{H}}+\sum\limits_{\zeta=1}%
^{m}\mathbf{p}_{\zeta}\times\overline{c}_{\text{Q,}\zeta},\text{
\ \ \ \ }\overline{r}_{\text{M}}=\mathbf{p}_{\text{H}}\times\overline
{r}_{\text{M,H}}+\sum\limits_{\zeta=1}^{m}\mathbf{p}_{\zeta}\times\overline
{r}_{\text{M,}\zeta},\\
\overline{r}_{\text{O}}  &  =\mathbf{p}_{\text{H}}\times\overline
{r}_{\text{O,H}}+\sum\limits_{\zeta=1}^{m}\mathbf{p}_{\zeta}\times\overline
{r}_{\text{O,}\zeta},\text{ \ \ \ }\overline{r}_{\text{U}}=\mathbf{p}%
_{\text{H}}\times\overline{r}_{\text{U,H}}+\sum\limits_{\zeta=1}^{m}%
\mathbf{p}_{\zeta}\times\overline{r}_{\text{U,}\zeta}.
\end{align*}
\end{The}

\textbf{Proof.} The proof is easy. We only take the first one as a example. To
do this, we need to consider two different cases:

\textbf{Case one: }If the the honest mining pool sets up the main chain in the
$r$th round of mining competition for $r=1,2,3,\ldots N$. In this case, we
have $c_{\text{Q,H}}^{(r)}=1$, and
\[
\frac{\sum\limits_{r=1}^{N_{\text{H}}}c_{\text{Q,H}}^{(r)}}{N}%
=\frac{N_{\text{H}}}{N}.
\]
This gives that as $N\rightarrow\infty$,%
\[
\frac{\sum\limits_{r=1}^{N_{\text{H}}}c_{\text{Q,H}}^{(r)}}{N}\rightarrow
\mathbf{p}_{\text{H}},\text{ \ a.s..}%
\]

\textbf{Case two: }If the $\zeta$th dishonest mining pool sets up the main
chain in the $r$th round of mining competition for $r=1,2,3,\ldots,N$. In this
case, it is worthwhile to note that $k_{\zeta}^{(r)}$ is the number of honest
blocks which have been mined by the honest mining pool before the $\zeta$th
dishonest mining pool begins to fork to a new sub-chain with $l_{\zeta
}^{\left(  r\right)  }$ blocks, and note that the part with $\varphi_{\zeta
}^{\left(  r\right)  }$ blocks of the main chain are pegged onto the blockchain,
thus we have%

\[
\ c_{\text{Q,}\zeta}^{\left(  r\right)  }=\frac{k_{\zeta}^{\left(  r\right)
}}{k_{\zeta}^{\left(  r\right)  }+\varphi_{\zeta}^{\left(  r\right)  }}.
\]

Since the competitively mining processes of the multiple mining pools are
repeated round after round, the random variable $c_{\text{Q,}\zeta}^{\left(
r\right)  }$ can be repeated by the similar experiments round after round
under the same conditions. Based on this, the random variable sequence
$\left\{  c_{\text{Q,}\zeta}^{\left(  r\right)  },r=1,2,3,\ldots\right\}  $
can be regarded as independent and identically distributed. When the number
that the $\zeta$th dishonest mining pool can set up the main chain is
$N_{\zeta}$, we obtain that as $N_{\zeta}\rightarrow\infty,$%
\[
\frac{\sum\limits_{r=1}^{N_{\zeta}}c_{\text{Q,}\zeta}^{\left(  r\right)  }%
}{N_{\zeta}}\rightarrow\overline{c}_{\text{Q,}\zeta},\ \text{a.s.},
\]
which holds for each $\zeta=1,2,\ldots,m$. This gives that as
$N\rightarrow\infty$,%

\[
\frac{\sum\limits_{r=1}^{N_{\zeta}}c_{\text{Q,}\zeta}^{\left(  r\right)  }}%
{N}=\frac{\sum\limits_{r=1}^{N_{\zeta}}c_{\text{Q,}\zeta}^{\left(  r\right)
}}{N_{\zeta}}\cdot\frac{N_{\zeta}}{N}=\mathbf{p}_{\zeta}\cdot\overline
{c}_{\text{Q,}\zeta},\text{ a.s.}.
\]

From the above two cases, we get%

\[
\overline{c}_{\text{Q}}=\mathbf{p}_{\text{H}}+\sum\limits_{\zeta=1}%
^{m}\mathbf{p}_{\zeta}\cdot\overline{c}_{\text{Q,}\zeta}.\text{ a.s..}%
\]

This completes the proof. $\square$

\section{The Growth Rate of Blockchain}

\label{Sec-7:growthrate}

In this section, we apply the renewal reward theory to study the long-term
growth rate of blockchain in the PoW Ethereum system with multiple mining pools.

In the $r$th round of mining competition, we denote by $v^{\left(  r\right)
}$ the number of blocks on the main chain when the honest mining pool sets up
the main chain, denote by $k_{i}^{\left(  r\right)  }$ and\ $l_{i}^{\left(
r\right)  }$ the number of honest blocks first mined by the honest mining pool
and the number of dishonest blocks mined by the $i$th dishonest mining pool
after forked when the $i$th dishonest mining pool sets up the main chain,
respectively. Note that the part with $\varphi_{i}^{\left(  r\right)  }$
blocks of the main chain is pegged onto the blockchain.

\begin{Lem}
In the PoW Ethereum system with multiple mining pools, by using the law of
large numbers, as $N\rightarrow\infty$, we have%
\[
\frac{\sum\limits_{r=1}^{N}v^{\left(  r\right)  }}{N}\rightarrow\overline
{v},\text{ \ a.s.,}%
\]
for $i=1,2,\ldots,m,$%
\[
\frac{\sum\limits_{r=1}^{N}k_{i}^{\left(  r\right)  }}{N}\rightarrow
\overline{k}_{i},\text{ \ a.s.,}%
\]%
\[
\frac{\sum\limits_{r=1}^{N}l_{i}^{\left(  r\right)  }}{N}\rightarrow
\overline{l}_{i},\text{ \ a.s.,}%
\]
and%
\[
\frac{\sum\limits_{r=1}^{N}\varphi_{i}^{\left(  r\right)  }}{N}\rightarrow
\overline{\varphi}_{i},\text{ \ a.s.,}%
\]
where $\overline{v}$, $\overline{k}_{i}$, $\overline{l}_{i}$ and
$\overline{\varphi}_{i}$ are the means of four random variables $v^{\left(
r\right)  }$, $k_{i}^{\left(  r\right)  },$ $l_{i}^{\left(  r\right)  }$\ and
$\varphi_{i}^{\left(  r\right)  }$, respectively.
\end{Lem}

It is worthwhile to note that in the PoW Ethereum system, the competitively
mining processes of the multiple mining pools are repeated round after round,
as we repeat the experiments under the same conditions. Thus, the moments that
one round of mining competition is over and the next round of mining
competition begins immediately are all renewal points. That is, let $T_{k}$ be
the $k$th moment that the $k$th round of mining competition is over and the
$\left(  k+1\right)  $th round of mining competition begins immediately. In
fact, $[T_{k-1},T_{k})$ represents a time interval that the $k$th round of
mining competition is underway. For simplicity of analysis, we assume that
$T_{0}=0$, i.e., the first round of mining competition begins at time
$T_{0}=0$. Let $N(t)=\max\{k,T_{k}\leq t\},$ then \{$N(t),t\geq0\}$ is a
renewal process.

We assume that there are $M_{k}$ blocks of the main chain in the time interval
$[T_{k-1},T_{k})$, and $M_{k}$ is independent of the time interval
$[T_{k-1},T_{k})$. Let $M\left(  t\right)  $ be the number of blocks of all
the main chains generated in the time interval $[0,t)$. Then, the following
theorem provides the growth rate $E\left[  M\left(  t\right)  \right]  /t$ of
blockchain in the PoW Ethereum system with multiple mining pools.

\begin{The}
In the PoW Ethereum system with multiple mining pools, if $E\left[
M_{1}\right]  <+\infty$ and $E\left[  T_{1}\right]  <+\infty$, then as
$t\rightarrow+\infty$, we have%
\begin{equation}
\frac{M\left(  t\right)  }{t}\rightarrow\frac{E\left[  M_{1}\right]
}{E\left[  T_{1}\right]  },\text{ a.s.,}\label{equ-1}%
\end{equation}%
\begin{equation}
\frac{E\left[  M\left(  t\right)  \right]  }{t}\rightarrow\frac{E\left[
M_{1}\right]  }{E\left[  T_{1}\right]  },\label{equ-2}%
\end{equation}
where%
\[
E\left[  M_{1}\right]  =\mathbf{p}_{\text{H}}\cdot\overline{v}+\sum_{i=1}%
^{m}\mathbf{p}_{i}\cdot\left\{  \overline{k}_{i}+\overline{\varphi}%
_{i}\right\}  .
\]
\end{The}

\textbf{Proof. }We give the proof for the equation \ref{equ-1} only. We write%

\[
\frac{M\left(  t\right)  }{t}=\frac{\sum\limits_{k=1}^{N(t)}M_{k}}%
{t}=\frac{\sum\limits_{k=1}^{N(t)}M_{k}}{N(t)}\cdot\frac{N(t)}{t}.
\]
Note that
\[
T_{k}=\left(  T_{1}-T_{0}\right)  +\left(  T_{2}-T_{1}\right)  +\cdots+\left(
T_{k}-T_{k-1}\right)
\]
is the time length of the $k$ renewal periods, and the random variables
$T_{1}-T_{0},$ $T_{2}-T_{1},\ldots,T_{k}-T_{k-1}$ are independent and
identically distributed, $E[T_{k}-T_{k-1}]=E[T_{1}-T_{0}]=E[T_{1}]$ for
$k\geq1$. At the same time, the random variables $M_{k},k\geq1$ are also
independent and identically distributed, $E[M_{k}]=E[M_{1}].$

Note that as $t\rightarrow+\infty,$ $N(t)\rightarrow+\infty.$ By the strong law
of large numbers, we obtain that as $t\rightarrow\infty$%

\[
\frac{\sum\limits_{k=1}^{N(t)}M_{k}}{N(t)}\rightarrow E[M_{1}].
\]

By using the elementary renewal theorem (Chapter 7 of \cite{Ross:2014}), we
obtain that%

\[
\frac{N(t)}{t}\rightarrow\frac{1}{E[T_{1}]},\text{ as }t\rightarrow+\infty.
\]

Therefore, we obtain that as $t\rightarrow+\infty$%

\[
\frac{M\left(  t\right)  }{t}\rightarrow\frac{E\left[  M_{1}\right]
}{E\left[  T_{1}\right]  }.
\]
This completes the proof. $\square$

\section{Average Reward Allocation among the Mining Pools}

\label{Sec-8:reward}

In this section, we propose a new method to compute the uncle block and nephew
block rewards in two consecutive rounds of mining competition, and provide
expressions for the long-term reward allocation and for the long-term reward
allocation rate to each mining pool by using the renewal reward theory.

we provide a computational method of the reward allocation among the honest
mining pool and the $m$th dishonest mining pools.

To set up the reward allocation, it is easy to see that a regular block is
paid the reward of $1$ block; an uncle block is paid the reward of $\left(
8-l\right)  /8$ blocks for $1\leq l\leq6$, where $l$ is the distance between
the uncle block and the nephew block; a nephew block is paid the reward of
$N_{\text{U}}/32$ blocks, where $N_{\text{U}}$ is the number of uncle blocks;
and a stale block is paid no reward.

In the PoW Ethereum system with multiple mining pools, it is worthwhile to
note that in the tree with multiple sub-chains, the main chain can be obtained
by either the honest mining pool or one of the $m$ dishonest mining pools,
thus our reward allocation is considered as the following two different cases.

\textbf{Case one: The honest mining pool sets up the main chain}

In this case, by observing (a) of Figure \ref{Fig-11}, we consider two
different cases as follows:

\textbf{(i)} The reward of the honest mining pool is given by%
\[
R_{\text{H,H}}=v+R_{\text{H,H}}^{\left(  \text{N}\right)  },
\]
where $v$ is the number of blocks in the main chain, $R_{\text{H,H}}^{\left(
\text{N}\right)  }$ is the reward of a nephew block which is the first block
of the main chain. Note that the nephew block refers to the uncle blocks in
the previous round of mining competition. Thus we obtain%
\[
R_{\text{H,H}}^{\left(  \text{N}\right)  }=\left\{
\begin{array}
[c]{ll}%
\frac{N_{\text{U,H}}}{32}, & \text{if the first block of the main chain is a
nephew block,}\\
0, & \text{otherwise.}%
\end{array}
\right.
\]

\textbf{(ii)} Note that each of the $m$ dishonest mining pools may mine only
the orphan blocks, that is, either the uncle blocks or the stale blocks. In
addition, it has a nephew block \textbf{either}\textit{ }if the main chain of
the previous round of mining competition is completitively pegged onto the
blockchain, and this dishonest mining pool is the first one to mine a block,
that is, the firstly mined block is the nephew block; \textbf{or} if this
dishonest mining pool sets up the main chain in the previous round of mining
competition, and a non-empty part of this main chain is left to this round of
mining competition, that is, the nephew block is the first block of the
non-empty part. Based on this, the reward of the $i$th dishonest mining pool
is given by%
\[
R_{\text{H},i}=R_{\text{H},i}^{\left(  \text{U}\right)  }+R_{\text{H}%
,i}^{(\text{N})},\text{ \ }i=1,2,\ldots,m,
\]
where $R_{\text{H},i}^{\left(  \text{U}\right)  }$ is the reward that if the
$i$th dishonest mining pool has an uncle block, and%
\[
R_{\text{H},i}^{\left(  \text{U}\right)  }=\frac{8-l}{8}%
\]
for $1\leq l\leq6$, where $l$ is the distance between the uncle block and the
nephew block of the next round of mining competition; and
\[
R_{\text{H},i}^{(\text{N})}=\frac{N_{\text{U,}i}}{32},
\]
where $N_{\text{U,}i}$ is the number of uncle blocks in the previous round of
mining competition if the nephew block belongs to the $i$th dishonest mining
pool. Based on this, we have%
\[
R_{\text{H},i}^{\left(  \text{U}\right)  }=\left\{
\begin{array}
[c]{ll}%
\frac{8-l}{8}, & \text{if the }i\text{th dishonest mining pool has an uncle
block,}\\
0, & \text{otherwise;}%
\end{array}
\right.
\]
and%
\[
R_{\text{H},i}^{(\text{N})}=\left\{
\begin{array}
[c]{ll}%
\frac{N_{\text{U},i}}{32}, & \text{if the nephew block belongs to the
}i\text{th dishonest mining pool,}\\
0, & \text{otherwise.}%
\end{array}
\right.
\]

\textbf{Case two: The }$\zeta$th\textbf{ dishonest mining pool sets up the
main chain}

In this case, by observing (b) and (c) of Figure \ref{Fig-11}, we consider
three different cases as follows:

\textbf{(i)} The reward of the $\zeta$th dishonest mining pool

Note that the $\zeta$th dishonest mining pool sets up the main chain, it is
clear that $\omega_{1}=k_{\zeta}+l_{\zeta}$. We assume that the part with
$\varphi_{\zeta}$ blocks for $\omega_{2}+2\leq\varphi_{\zeta}\leq\omega_{1}$
of the main chain are pegged onto the blockchain, while another part with
$\omega_{1}-\varphi_{\zeta}$ blocks of the main chain is left to the next
round of mining competition.

The reward of the $\zeta$th dishonest mining pool is given by%
\[
R_{\zeta,\zeta}=\varphi_{\zeta}+R_{\zeta,\zeta}^{(\text{N})},
\]
where $\varphi_{\zeta}$ is the number of blocks in the part of the main chain,
which is pegged onto the blockchain; and
\[
R_{\zeta,\zeta}^{(\text{N})}=\frac{N_{\text{U},\zeta}}{32},
\]
where $N_{\text{U},\zeta}$ is the number of uncle blocks in the previous round
of mining competition if the nephew block belongs to the $\zeta$th dishonest
mining pool. Based on this, we have%
\[
R_{\zeta,\zeta}^{(\text{N})}=\left\{
\begin{array}
[c]{ll}%
\frac{N_{\text{U},\zeta}}{32}, & \text{if the nephew block belongs to the
}\zeta\text{th dishonest mining pool,}\\
0, & \text{otherwise.}%
\end{array}
\right.
\]

\textbf{(ii)} The reward of the honest mining pool

If the $\zeta$th dishonest mining pool sets up the main chain, then the honest
mining pool can mine only the orphan blocks after the $\zeta$th dishonest
mining pool forks, that is, either the uncle blocks or the stale blocks. In
addition, it may have a nephew block if the honest mining pool is the first
one to mine a block in this round of mining competition. In this case, the
reward of the honest mining pool contains the reward of $k_{\zeta}$ regular
blocks, the reward of one uncle block and the reward of a nephew block. Based
on this, the reward of the honest mining pool is given by%
\[
R_{\zeta,\text{H}}=k_{\zeta}+R_{\zeta,\text{H}}^{\left(  \text{U}\right)
}+R_{\zeta,\text{H}}^{\left(  \text{N}\right)  },
\]
where $R_{\zeta,\text{H}}^{\left(  \text{U}\right)  }$ is the reward that if
the honest mining pool has an uncle block, and%
\[
R_{\zeta,\text{H}}^{\left(  \text{U}\right)  }=\frac{8-l}{8}%
\]
for $1\leq l\leq6$, where $l$ is the distance between the uncle block and the
nephew block of the next round of mining competition; and
\[
R_{\zeta,\text{H}}^{\left(  \text{N}\right)  }=\frac{N_{\text{U,H}}}{32},
\]
where $N_{\text{U,H}}$ is the number of uncle blocks in the previous round of
mining competition if the nephew block belongs to the honest mining pool.
Based on this, we have%
\[
R_{\zeta,\text{H}}^{\left(  \text{U}\right)  }=\left\{
\begin{array}
[c]{ll}%
\frac{8-l}{8}, & \text{if the honest mining pool has an uncle block,}\\
0, & \text{otherwise;}%
\end{array}
\right.
\]
and%
\[
R_{\zeta,\text{H}}^{\left(  \text{N}\right)  }=\left\{
\begin{array}
[c]{ll}%
\frac{N_{\text{U,H}}}{32}, & \text{if the nephew block belongs to the honest
mining pool,}\\
0, & \text{otherwise.}%
\end{array}
\right.
\]

\textbf{(iii)} The reward of the $i$th dishonest mining pool for $i\neq\zeta$

If the $\zeta$th dishonest mining pool sets up the main chain, then the $i$th
dishonest mining pool can mine only the orphan blocks, that is, either the
uncle blocks or the stale blocks. In addition, it has a nephew block
\textbf{either}\textit{ }if the main chain of the previous round of mining
competition is completely pegged onto the blockchain, and the $i$th dishonest
mining pool is the first one to mine a block, that is, the firstly mined block
is the nephew block that belongs to the $i$th dishonest mining pool; \textbf{or} if
the $i$th dishonest mining pool sets up the main chain in the previous round
of mining competition, and a non-empty part of this main chain is left to this
round of mining competition, that is, the nephew block is the first block of
the non-empty part. Based on this, the reward of the $i$th dishonest mining
pool is given by%
\[
R_{\zeta,i}=R_{\zeta,i}^{\left(  \text{U}\right)  }+R_{\zeta,i}^{\left(
\text{N}\right)  },\text{ \ }i=1,2,\ldots,m,
\]
where $R_{\zeta,i}^{\left(  \text{U}\right)  }$ is the reward that if the
$i$th dishonest mining pool has an uncle block, and%
\[
R_{\zeta,i}^{\left(  \text{U}\right)  }=\frac{8-l}{8}%
\]
for $1\leq l\leq6$, where $l$ is the distance between the uncle block and the
nephew block of the next round of mining competition; and
\[
R_{\zeta,i}^{\left(  \text{N}\right)  }=\frac{N_{\text{U},i}}{32},
\]
where $N_{\text{U},i}$ is the number of uncle blocks in the previous round of
mining competition if the nephew block belongs to the $i$th dishonest mining
pool. Based on this, we have%
\[
R_{\zeta,i}^{\left(  \text{U}\right)  }=\left\{
\begin{array}
[c]{ll}%
\frac{8-l}{8}, & \text{if the }i\text{th dishonest mining pool has an uncle
block,}\\
0, & \text{otherwise;}%
\end{array}
\right.
\]
and%
\[
R_{\zeta,i}^{\left(  \text{N}\right)  }=\left\{
\begin{array}
[c]{ll}%
\frac{N_{\text{U},i}}{32}, & \text{if the nephew block belongs to the
}i\text{th dishonest mining pool,}\\
0, & \text{otherwise.}%
\end{array}
\right.  .
\]

When the reward of each mining pool is regarded as random variables, we
can apply the law of large numbers to further study the reward of each mining
pool. In this situation, we obtain some interesting results, which can be
applied to solving many practical problems owing to the fact that by using the
law of large numbers, our experimental reward of each mining pool can steadily
approach their corresponding fixed values. Such a statistical method is
effective and useful in the study of the PoW Ethereum system with multiple
mining pools, because analysis of the tree with multiple sub-chains always has
a higher computational complexity.

In what follows, we apply the law of large numbers to study the reward
obtained by each multiple mining pool in the PoW Ethereum system. Let
$R_{\text{H}}^{(r)}$ and $R_{i}^{(r)}$ be the reward obtained by the honest
mining pool or the $i$th dishonest mining pool in the $r$th round of
competition for $i=1,2,\ldots,m$.

\begin{The}
In the PoW Ethereum system with multiple mining pools, by using the law of
large numbers, as $N\rightarrow\infty$, we have%
\[
\frac{\sum\limits_{r=1}^{N}R_{\text{H}}^{(r)}}{N}\rightarrow\overline
{R}_{\text{H}},\text{ \ a.s.,}%
\]
and for $i=1,2,\ldots,m,$%
\[
\frac{\sum\limits_{r=1}^{N}R_{i}^{(r)}}{N}\rightarrow\overline{R}_{i},\text{
\ a.s..}%
\]
\end{The}

\textbf{Proof.} The proof is easy. We only take the first one as a proof
example. Here, we also need to consider two different cases:

\textbf{Case one: }\textbf{ }The honest mining pool sets up the main
chain in the $r$th round of mining competition for $r=1,2,3,\ldots,N$. In this
case,
\[
R_{\text{H,H}}^{(r)}=v^{\left(  r\right)  }+R_{\text{H,H}}^{\left(
\text{N},r\right)  },
\]
where in the $r$th round of mining competition, $v^{\left(  r\right)  }$ and
$R_{\text{H}}^{\left(  \text{N},r\right)  }$ are the number of blocks in the
main chain, and the reward of a nephew block which is the first block of the
main chain, respectively. Note that the main chain is mined by the honest
mining pool.

Note that the competitively mining processes of the multiple mining pools are
repeated round after round, the random variables $v^{\left(  r\right)  }$ and
$R_{\text{H}}^{\left(  \text{N},r\right)  }$ (thus $R_{\text{H}}^{(r)}$) can
be repeated by the experiments round after round under the same conditions.
Based on this, the random variable sequences $\left\{  v^{\left(  r\right)
},r=1,2,3,\ldots\right\}  $ and $\left\{  R_{\text{H,H}}^{\left(
\text{N},r\right)  },r=1,2,3,\ldots\right\}  $ (thus $\left\{  R_{\text{H}%
}^{(r)},r=1,2,3,\ldots\right\}  $) can be regarded as independent and
identically distributed. Therefore, we obtain%
\[
\frac{\sum\limits_{r=1}^{N_{\text{H}}}v^{\left(  r\right)  }}{N_{\text{H}}%
}\rightarrow\overline{v},\text{ \ a.s.,}%
\]
and%
\[
\frac{\sum\limits_{r=1}^{N_{\text{H}}}R_{\text{H,H}}^{\left(  \text{N}%
,r\right)  }}{N_{\text{H}}}\rightarrow\overline{R}_{\text{H,H}}^{\left(
\text{N}\right)  },\text{ \ a.s..}%
\]
This gives%
\begin{align*}
\frac{\sum\limits_{r=1}^{N_{\text{H}}}R_{\text{H,H}}^{(r)}}{N_{\text{H}}}  &
=\frac{\sum\limits_{r=1}^{N_{\text{H}}}v^{\left(  r\right)  }}{N_{\text{H}}%
}+\frac{\sum\limits_{r=1}^{N_{\text{H}}}R_{\text{H,H}}^{\left(  \text{N}%
,r\right)  }}{N_{\text{H}}}\\
&  \rightarrow\overline{v}+\overline{R}_{\text{H,H}}^{\left(  \text{N}\right)
},\text{ \ \ \ \ \ \ \ \ a.s..}%
\end{align*}

\textbf{Case two: }The $\zeta$th dishonest mining pool sets up the main
chain in the $r$th round of mining competition for $r=1,2,3,\ldots$.

In this case, the honest mining pool can mine only the orphan blocks, that is,
either the uncle blocks or the stale blocks. In addition, it may have a nephew
block if the honest mining pool is the first one to mine a block in a round of
mining competition, that is, the firstly mined block is the nephew block.
Thus, we have%
\[
R_{\zeta,\text{H}}^{(r)}=k_{\zeta}^{(r)}+R_{\zeta,\text{H}}^{\left(
\text{U},r\right)  }+R_{\zeta,\text{H}}^{\left(  \text{N},r\right)  },
\]
where in the $r$th round of mining competition, $R_{\zeta,\text{H}}^{\left(
\text{U},r\right)  }$ is the reward of $1$ uncle block if the honest mining
pool has an uncle block.

Note that the competitively mining processes of the multiple mining pools are
repeated round after round, the random variables $k_{\zeta}^{(r)},$
$R_{\zeta,\text{H}}^{\left(  \text{U},r\right)  }$ and $R_{\zeta,\text{H}%
}^{\left(  \text{N},r\right)  }$ (thus $R_{\zeta,\text{H}}^{(r)}$) can be
repeated by the experiments round after round under the same conditions. Based
on this, the random variable sequences $\left\{  k_{\zeta}^{(r)}%
,r=1,2,3,\ldots\right\}  ,$ $\left\{  R_{\zeta,\text{H}}^{\left(
\text{U},r\right)  },r=1,2,3,\ldots\right\}  $ and $\left\{  R_{\zeta
,\text{H}}^{\left(  \text{N},r\right)  },r=1,2,3,\ldots\right\}  $ (thus
$\left\{  R_{\zeta,\text{H}}^{(r)},r=1,2,3,\ldots\right\}  $) can be regarded
as independent and identically distributed. Therefore, we obtain%

\[
\frac{\sum\limits_{r=1}^{N_{\zeta}}k_{\zeta}^{(r)}}{N_{\zeta}}\rightarrow
\overline{k}_{\zeta},\text{ \ a.s.,}%
\]%

\[
\frac{\sum\limits_{r=1}^{N_{\zeta}}R_{\zeta,\text{H}}^{\left(  \text{U}%
,r\right)  }}{N_{\zeta}}\rightarrow\overline{R}_{\zeta,\text{H}}^{\left(
\text{U}\right)  },\text{ \ a.s.,}%
\]
and%
\[
\frac{\sum\limits_{r=1}^{N_{\zeta}}R_{\zeta,\text{H}}^{\left(  \text{N}%
,r\right)  }}{N_{\zeta}}\rightarrow\overline{R}_{\zeta,\text{H}}^{\left(
\text{N}\right)  },\text{ \ a.s..}%
\]
This gives%
\begin{align*}
\frac{\sum\limits_{r=1}^{N_{\zeta}}R_{\zeta,\text{H}}^{(r)}}{N_{\zeta}}  &
=\frac{\sum\limits_{r=1}^{N_{\zeta}}k_{\zeta}^{(r)}}{N_{\zeta}}+\frac{\sum
\limits_{r=1}^{N_{\zeta}}R_{\zeta,\text{H}}^{\left(  \text{U},r\right)  }%
}{N_{\zeta}}+\frac{\sum\limits_{r=1}^{N_{\zeta}}R_{\zeta,\text{H}}^{\left(
\text{N},r\right)  }}{N_{\zeta}}\\
&  \rightarrow\overline{k}_{\zeta}+\overline{R}_{\zeta,\text{H}}^{\left(
\text{U}\right)  }+\overline{R}_{\zeta,\text{H}}^{\left(  \text{N}\right)
},\text{ \ \ \ \ \ \ \ \ a.s..}%
\end{align*}
Therefore, we can get that as $N\rightarrow\infty$%

\begin{align*}
\frac{\sum\limits_{r=1}^{N}R_{\text{H}}^{(r)}}{N}  &  =\frac{\sum
\limits_{r=1}^{N_{\text{H}}}R_{\text{H}}^{(r)}}{N}+\frac{\sum\limits_{\zeta
=1}^{m}\sum\limits_{r=1}^{N_{\zeta}}R_{\zeta,\text{H}}^{(r)}}{N}\\
&  =\frac{\sum\limits_{r=1}^{N_{\text{H}}}R_{\text{H}}^{(r)}}{N_{\text{H}}%
}\cdot\frac{N_{\text{H}}}{N}+\frac{\sum\limits_{\zeta=1}^{m}\sum
\limits_{r=1}^{N_{\zeta}}R_{\zeta,\text{H}}^{(r)}}{N_{\zeta}}\cdot
\frac{N_{\zeta}}{N}\\
&  =\mathbf{p}_{\text{H}}\left(  \overline{v}+\overline{R}_{\text{H,H}%
}^{\left(  \text{N}\right)  }\right)  +\sum\limits_{\zeta=1}^{m}%
\mathbf{p}_{\zeta}\left(  \overline{k}_{\zeta}+\overline{R}_{\zeta,\text{H}%
}^{\left(  \text{U}\right)  }+\overline{R}_{\zeta,\text{H}}^{\left(
\text{N}\right)  }\right)  ,\text{ a.s..}%
\end{align*}
This completes the proof. $\square$

\section{Reward Rates Allocated among the Mining Pools}

\label{Sec-9:rewardrate}

In this section, we apply the renewal reward processes to study the long-term
reward allocation rates among the multiple mining pools.

Note that the competitively mining processes of the multiple mining pools are
repeated round after round, as we repeat the experiments under the same
conditions. Thus, the moments that one round of mining competition is over and
the next round of mining competition begins immediately are all renewal
points. Let $T_{k}$ be the $k$th moment that the $k$th round of mining
competition is over and the $\left(  k+1\right)  $th round of mining
competition begins immediately. In fact, $[T_{k-1},T_{k})$ represents such a
time interval that the $k$th round of mining competition is underway. For
simplicity, we assume that $T_{0}=0$. Let $N(t)=\max\{k,T_{k}\leq
t\},$ then \{$N(t),t\geq0\}$ is a renewal process.

Let $R_{\text{H}}^{\left(  k\right)  }$ and $R_{\zeta}^{\left(  k\right)  }%
$\ be the rewards allocated to the honest mining pool and the $\zeta$th
dishonest mining pool in the time interval $[T_{k-1},T_{k})$, respectively.
Meanwhile, $R_{\text{H}}^{\left(  k\right)  }$ and $R_{\zeta}^{\left(
k\right)  }$ are independent of the time interval $[T_{k-1},T_{k})$. Let
$R_{\text{H}}\left(  t\right)  $ and $R_{\zeta}\left(  t\right)  $\ be the
rewards allocated to the honest mining pool and the $\zeta$th dishonest mining
pool in the time interval $[0,t)$. The following theorem respectively provides
the reward allocation rates to the honest mining pool and the $\zeta$th
dishonest mining pool in the PoW Ethereum system with multiple mining pools.

\begin{The}
In the PoW Ethereum system with multiple mining pools, if $E\left[
R_{\text{H}}^{\left(  1\right)  }\right]  <+\infty$ and $E\left[
T_{1}\right]  <+\infty$, then as $t\rightarrow\infty$,%
\begin{equation}
\frac{R_{\text{H}}\left(  t\right)  }{t}\rightarrow\frac{E\left[  R_{\text{H}%
}^{\left(  1\right)  }\right]  }{E\left[  T_{1}\right]  },\text{
\ a.s.,}\label{equ-3}%
\end{equation}%
\begin{equation}
\frac{E\left[  R_{\text{H}}\left(  t\right)  \right]  }{t}\rightarrow
\frac{E\left[  R_{\text{H}}^{\left(  1\right)  }\right]  }{E\left[
T_{1}\right]  };\label{equ-4}%
\end{equation}%
\begin{equation}
\frac{R_{\zeta}\left(  t\right)  }{t}\rightarrow\frac{E\left[  R_{\zeta
}^{\left(  1\right)  }\right]  }{E\left[  T_{1}\right]  },\text{
\ a.s.,}\label{equ-5}%
\end{equation}%
\begin{equation}
\frac{E\left[  R_{\zeta}\left(  t\right)  \right]  }{t}\rightarrow
\frac{E\left[  R_{\zeta}^{\left(  1\right)  }\right]  }{E\left[  T_{1}\right]
},\label{equ-6}%
\end{equation}
where%
\begin{align*}
E\left[  R_{\text{H}}^{\left(  1\right)  }\right]  = &  \mathbf{p}_{\text{H}%
}\left(  \overline{v}+\mathbf{q}_{\text{H},\text{H}}^{\left(  \text{N}\right)
}\overline{R}_{\text{H,H}}^{\left(  \text{N}\right)  }\right)  \\
&  +\sum_{\zeta=1}^{m}\mathbf{p}_{\zeta}\left(  \overline{k}_{\zeta
}+\mathbf{q}_{\zeta,\text{H}}^{\left(  \text{U}\right)  }\overline{R}%
_{\zeta,\text{H}}^{\left(  \text{U}\right)  }+\mathbf{q}_{\zeta,\text{H}%
}^{\left(  \text{N}\right)  }\overline{R}_{\zeta,\text{H}}^{\left(
\text{N}\right)  }\right)  ,
\end{align*}%
\begin{align*}
E\left[  R_{\zeta}^{\left(  1\right)  }\right]  = &  \mathbf{p}_{\zeta}\left(
\overline{\varphi}_{\zeta}+\mathbf{q}_{\zeta,\zeta}^{\left(  \text{N}\right)
}\overline{R}_{\zeta,\zeta}^{\left(  \text{N}\right)  }\right)  \\
&  +\mathbf{p}_{\text{H}}\left(  \mathbf{q}_{\text{H},\zeta}^{\left(
\text{U}\right)  }\overline{R}_{\text{H},\zeta}^{\left(  \text{U}\right)
}+\mathbf{q}_{\text{H},\zeta}^{\left(  \text{N}\right)  }\overline
{R}_{\text{H},\zeta}^{\left(  \text{N}\right)  }\right)  \\
&  +\sum_{k\neq\zeta}^{m}\mathbf{p}_{k}\left(  \mathbf{q}_{k,\zeta}^{\left(
\text{U}\right)  }\overline{R}_{\text{k},\zeta}^{\left(  \text{U}\right)
}+\mathbf{q}_{k,\zeta}^{\left(  \text{N}\right)  }\overline{R}_{\text{k}%
,\zeta}^{\left(  \text{N}\right)  }\right)  .
\end{align*}
\end{The}

\textbf{Proof. }We give the proof for equation \ref{equ-3} only. To do
this, we write%

\[
\frac{R_{\text{H}}\left(  t\right)  }{t}=\frac{\sum\limits_{k=1}%
^{N(t)}R_{\text{H}}^{\left(  k\right)  }}{t}=\frac{\sum\limits_{k=1}%
^{N(t)}R_{\text{H}}^{\left(  k\right)  }}{N(t)}\cdot\frac{N(t)}{t}.
\]
Note that
\[
T_{k}=\left(  T_{1}-T_{0}\right)  +\left(  T_{2}-T_{1}\right)  +\cdots+\left(
T_{k}-T_{k-1}\right)
\]
is the time length of the $k$ renewal periods, and the random variables
$T_{1}-T_{0},$ $T_{2}-T_{1},\ldots,T_{k}-T_{k-1}$ are independent and
identically distributed, $E[T_{k}-T_{k-1}]=E[T_{1}-T_{0}]=E[T_{1}]$ for
$k\geq1$. At the same time, the random variables $R_{\text{H}}^{\left(
k\right)  },k\geq1$ are also independent and identically distributed,
$E[R_{\text{H}}^{\left(  k\right)  }]=E[R_{\text{H}}^{\left(  1\right)  }].$

Note that $t\rightarrow+\infty,$ $N(t)\rightarrow+\infty.$ By the strong law
of large numbers, we obtain that as $t\rightarrow\infty$%

\[
\frac{\sum\limits_{k=1}^{N(t)}R_{\text{H}}^{\left(  k\right)  }}%
{N(t)}\rightarrow E[R_{\text{H}}^{\left(  1\right)  }].
\]

According to the elementary renewal theorem (Chapter 7 of \cite{Ross:2014}),
we obtain that%

\[
\frac{N(t)}{t}\rightarrow\frac{1}{E[T_{1}]},\text{ as }t\rightarrow+\infty.
\]

Therefore, we obtain that as $t\rightarrow+\infty,$%

\[
\frac{R_{\text{H}}\left(  t\right)  }{t}\rightarrow\frac{E\left[  R_{\text{H}%
}^{\left(  1\right)  }\right]  }{E\left[  T_{1}\right]  }.
\]
This completes the proof. $\square$

\section{Simulation Experiments}

\label{Sec-10:simulation}

In this section, we use some simulation experiments to discuss the Ethereum
system with one honest mining pool and two dishonest mining pools, verify how
the key probabilities of Ethereum are obtained approximately by using the law
of large numbers, and analyze the performance measures of the Ethereum system
by means of the renewal reward theorem.

\subsection{An Experiment Design}

In order to analyze the competitive mining processes of multiple mining pools,
our simulation experiment is designed as follows:

\textbf{(1)} In our simulation, we take a round of mining competition as a
sampling, and then we repeat such a sampling round after round.

\textbf{(2)} Under the two-block leading competitive criterion, one dishonest
mining pool may release a part of its sub-chain into the Ethereum system.
Here, our simulation experiments consider the mining rule: $\omega_{1}%
-\omega_{2}=2$. That is, the longest sub-chain length is 2 blocks longer than
the second-longest sub-chain in the system.

Although $\omega_{1}-\omega_{2}=2$ is taken as a terminate rule to end a round
of mining competition among the multiple mining pools, the dishonest mining
pools can fork at any position of the sub-chain mined by the honest mining
pool. Thus, the tree with multiple block sub-chains is still more complicated,
and also represents the general practical mining structure of the PoW Ethereum
system with multiple mining pools.

\textbf{(3)} For simplicity of simulation, our design is to consider the
mining processes of the multiple mining pools. To do this, we take $m+1$
random variables as
\[
X_{i}=X\cdot(\frac{1}{\alpha_{i}}+\frac{1}{\gamma}),i=0,1,2,...,m,
\]
where, $X_{i}$ is the block-generating and block-pegging time of the $i$th
mining pool, $\alpha_{i}$ is the mining power of the $i$th mining pool, and
$\gamma$ is the communication ability of the P2P network. Note that
$1/\alpha_{i}+1/\gamma$ is used to show the independence between the mining
power and the communication ability. At the same time, it is easy to see that
$X_{i}$ decreases as $\alpha_{i}$ or $\gamma$ increases, this is consistent
with our intuitive understanding on the mining times.

We assume that the random variable $X$ obeys an exponential distribution of
the mean $15$ seconds, where the $15$ seconds are always chosen as the
expected mining time of one block in the PoW Ethereum system. Let $d$ be a
random number generated by the exponential distribution of $X$, then we have%

\[
d_{i}=d\cdot(\frac{1}{\alpha_{i}}+\frac{1}{\gamma}),i=0,1,2,...,m.
\]

\subsection{Simulation and results}

In this subsection, we describe and analyze some interesting simulation results.

\textbf{(1) The law of large numbers}

Note that the probability that each mining pool wins the mining competition
plays a key role in our research on the key ratios, the growth rate of
blockchain, the reward allocation rates and so on. Here, it is necessary to
verify how this probability is obtained approximately by using the law of
large numbers. To this end, we take $N\in\left[  10000,40000\right]  $
and $\gamma=10.$

\begin{figure}[ptbh]
\centering                                       \subfigure[$\alpha_{\text{H}%
}=0.6,\alpha_{1}=0.3,\alpha_{2}%
=0.1$ ]{\includegraphics[width=7cm]{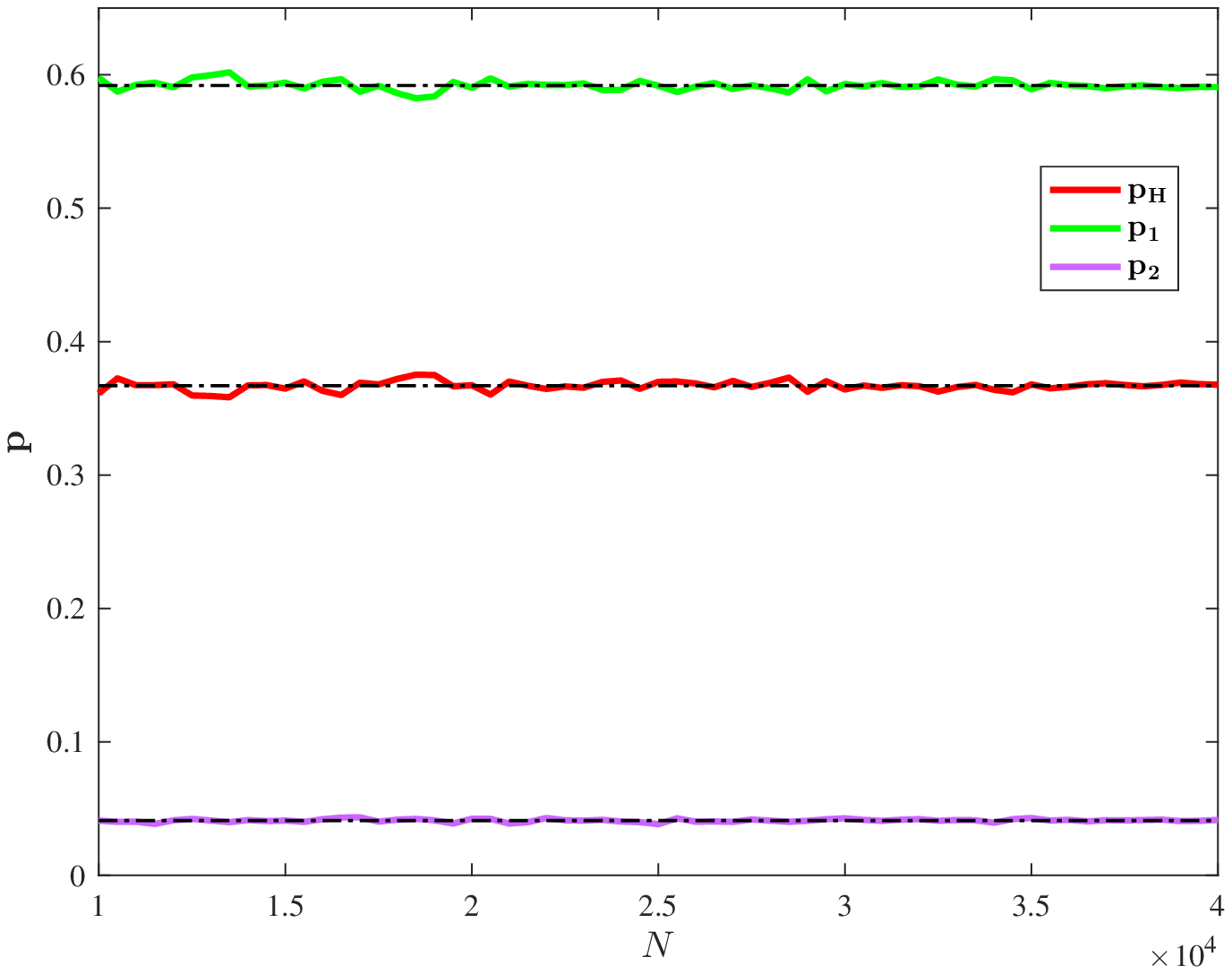}}
\subfigure[$\alpha_{\text{H}}=0.7,\alpha_{1}=0.2,\alpha_{2}%
=0.1$] { \includegraphics[width=7cm]{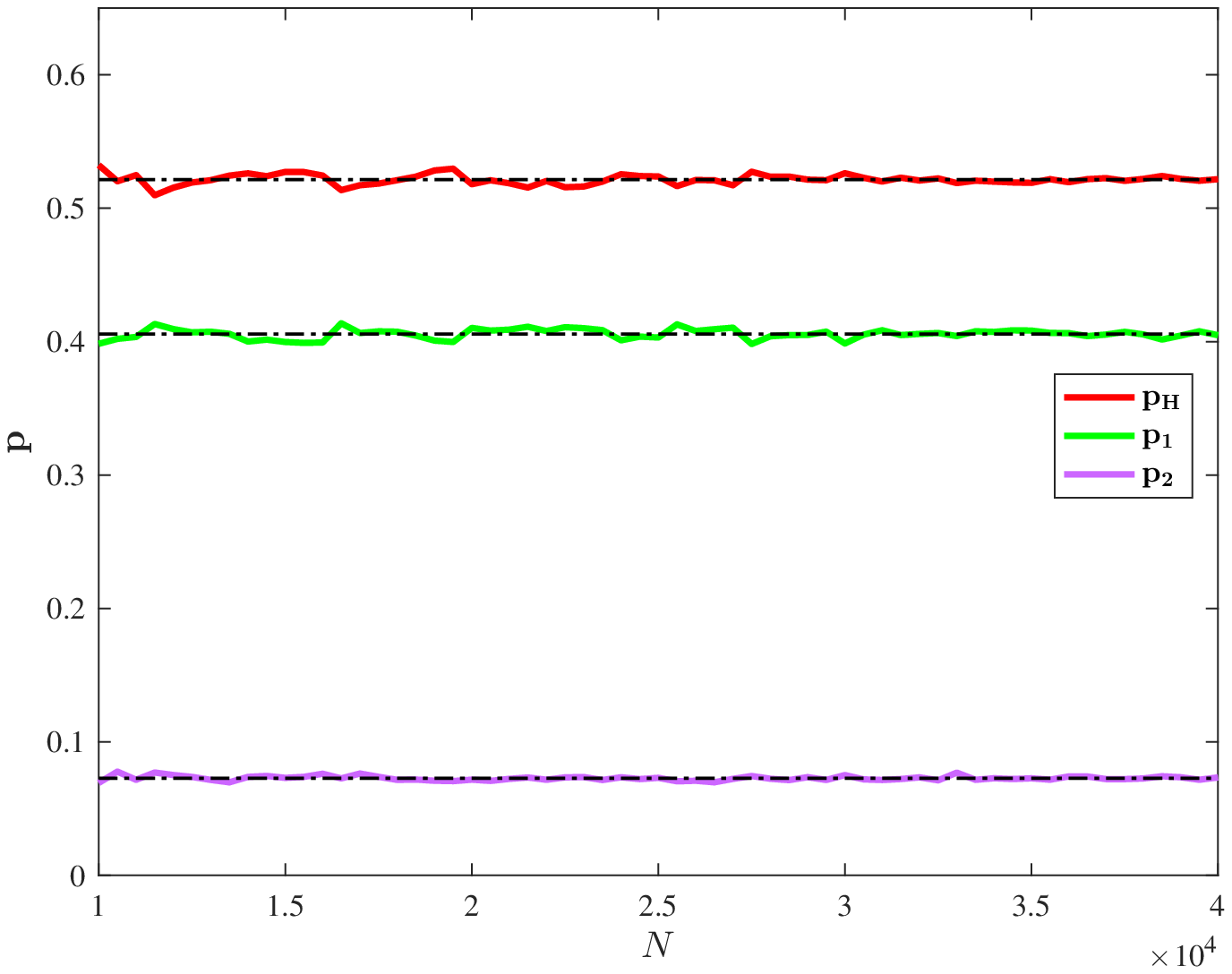}}  \caption{The
probabilities are approximately computed by the law of large numbers.}%
\label{Fig-12}%
\end{figure}

From Figure \ref{Fig-12}, it is easy to see that the three probabilities:
$\mathbf{p}_{\text{H}},$ $\mathbf{p}_{1},$and $\mathbf{p}_{2}$ fluctuate
around a certain value. This shows that the laws of large numbers is well
applied to determine these probabilities.

From the left part of Figure \ref{Fig-12}, it is seen that $\mathbf{p}%
_{1}>\mathbf{p}_{\text{H}},$ while $\mathbf{p}_{1}<\mathbf{p}_{\text{H}}$ in
the right of the Figure \ref{Fig-12}. It shows that as $\alpha_{\text{H}}$
increases, there exists a $\alpha_{\text{H}}^{\ast},$ such that $\mathbf{p}%
_{1}=\mathbf{p}_{\text{H}}.$ Furthermore, it is observed that as
$\alpha_{\text{H}}$ increases, the probability $\mathbf{p}_{1}+\mathbf{p}_{2}$
of the dishonest mining pools decreases. Thus, the influence of the dishonest
pools decreases as $\alpha_{\text{H}}$ increases.

\begin{figure}[ptbh]
\centering                 \includegraphics[width=10cm]{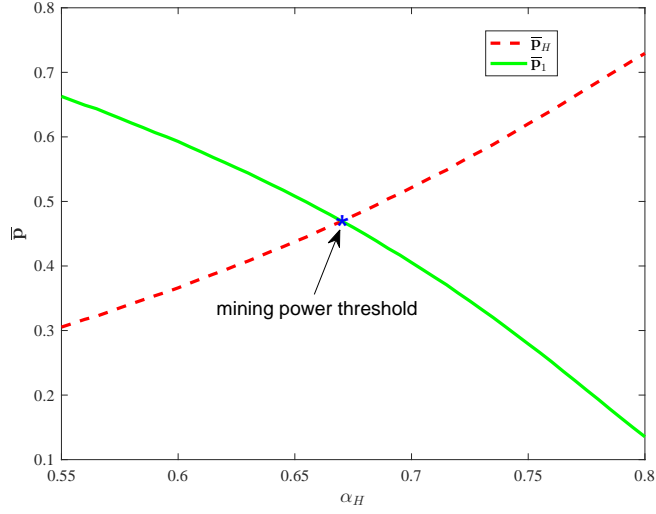}
\centering   \caption{The average probabilities and the mining power
threshold.}%
\label{Fig-13}%
\end{figure}

In order to observe the interesting value $\alpha_{\text{H}}^{\ast},$ we use a
special experiment. Let $N=20000,\gamma=10,\alpha_{2}=0.1,$ $\alpha_{\text{H}%
}\in\left[  0.55,0.8\right]  ,$ $\alpha_{1}=1-\alpha_{\text{H}}-\alpha_{2}.$
Each of our simulations with $20000$ rounds of mining competition is repeated
$100$ times to calculate the average of the approximate probabilities
$\mathbf{p}_{\text{H}}$ and $\mathbf{p}_{1}.$ We denote the two average values
by $\overline{\mathbf{p}}_{\text{H}}$ and $\overline{\mathbf{p}}_{1}$. The
results are shown in Figure \ref{Fig-13}.

From Figure \ref{Fig-13}, it is seen that as $\alpha_{\text{H}}$ increases,
the average probability $\overline{\mathbf{p}}_{\text{H}}$ increases, and the
average probability $\overline{\mathbf{p}}_{1}$ decreases. Also, it is
observed that there exists a $\alpha_{\text{H}}^{\ast},$ when $\alpha
_{\text{H}}<\alpha_{\text{H}}^{\ast}$, $\overline{\mathbf{p}}_{1}%
>\overline{\mathbf{p}}_{\text{H}}$; and when $\alpha_{\text{H}}>\alpha
_{\text{H}}^{\ast}$, $\overline{\mathbf{p}}_{1}<\overline{\mathbf{p}%
}_{\text{H}}$. In our simulation experiments, we obtain that\ the $95\%$
confidence interval of the mining power $\alpha_{\text{H}}^{\ast}$ is $\left[
0.6662,0.6731\right]  $. This shows that when the mining power of the $1$st
dishonest mining pool exceeds the mining power threshold $23.38\%$, the $1$st
dishonest pool has the biggest probability of setting up the main chain. Based
on this, the dishonest mining pool with a smaller mining power can have the
same probability of setting up the main chain as the honest mining pool. This
is why the dishonest mining pool may fork at any position of the honest sub-chain.

In the remainder of this subsection, we explore how the performance measures
of the Ethereum system depend on the mining powers of the three mining pools.
To this end, we take the parameters as follows: The mining power of the honest
mining pool $\alpha_{\text{H}}\in\left[  0.51,0.82\right]  ,$ the mining power
of the $2$nd dishonest mining pool $\alpha_{2}=0.13,$ and the mining power of
the $1$st dishonest mining pool $\alpha_{1}=1-\alpha_{\text{H}}-\alpha_{2},$
the rate at which the block is pegged to the corresponding sub-chain
$\gamma=10$, and the numbers of mining competition rounds are $N_{1}=15000,$
$N_{2}=18000,N_{3}=20000,N_{4}=25000,N_{5}=30000$.

\textbf{ (2) Some key probabilities vs. $\alpha_{\text{H}}$}

\begin{figure}[ptbh]
\centering                          \subfigure[]{
\begin{minipage}%
[]{0.4\linewidth}
\includegraphics[width=1\linewidth]{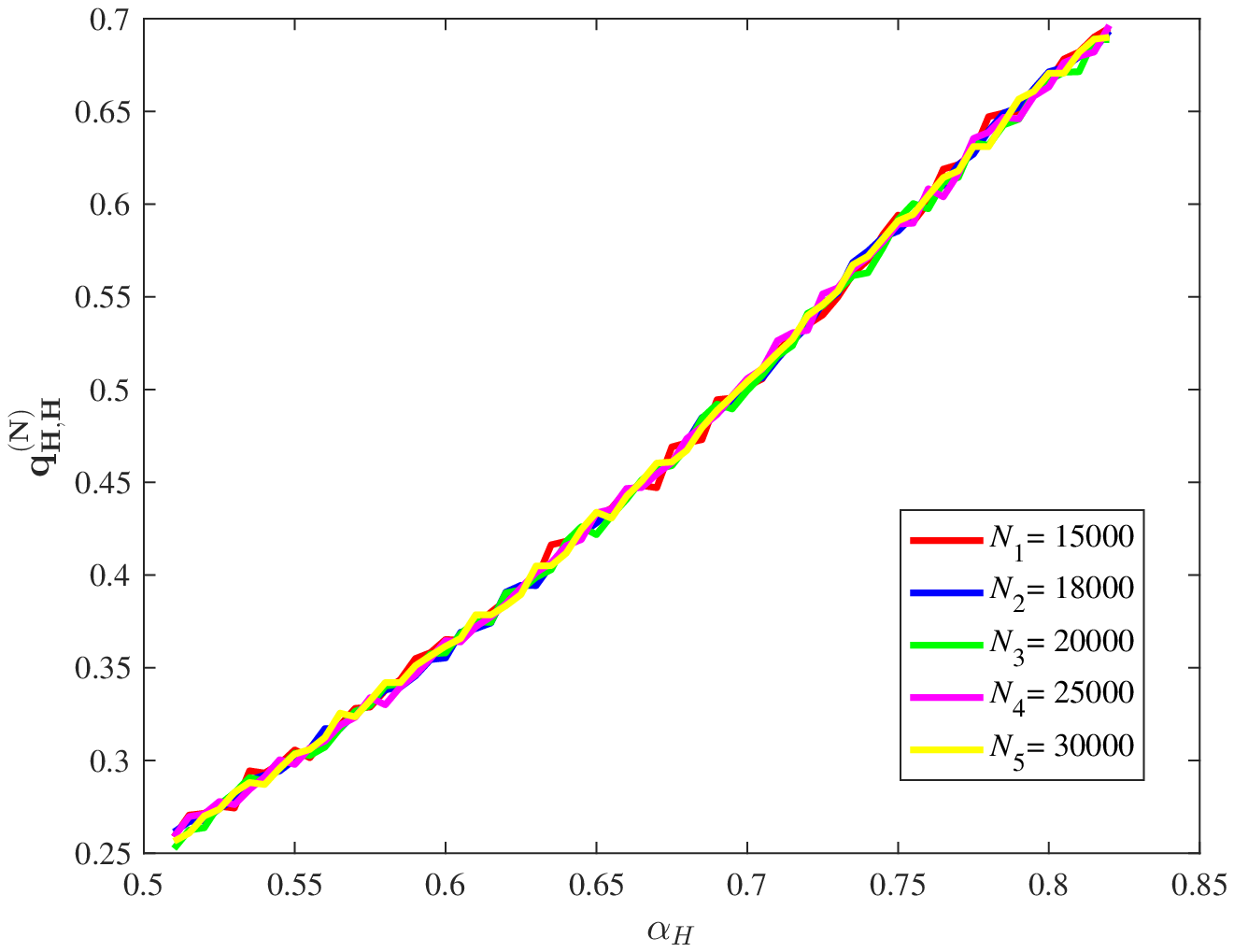}%

\
 \includegraphics[width=1\linewidth]{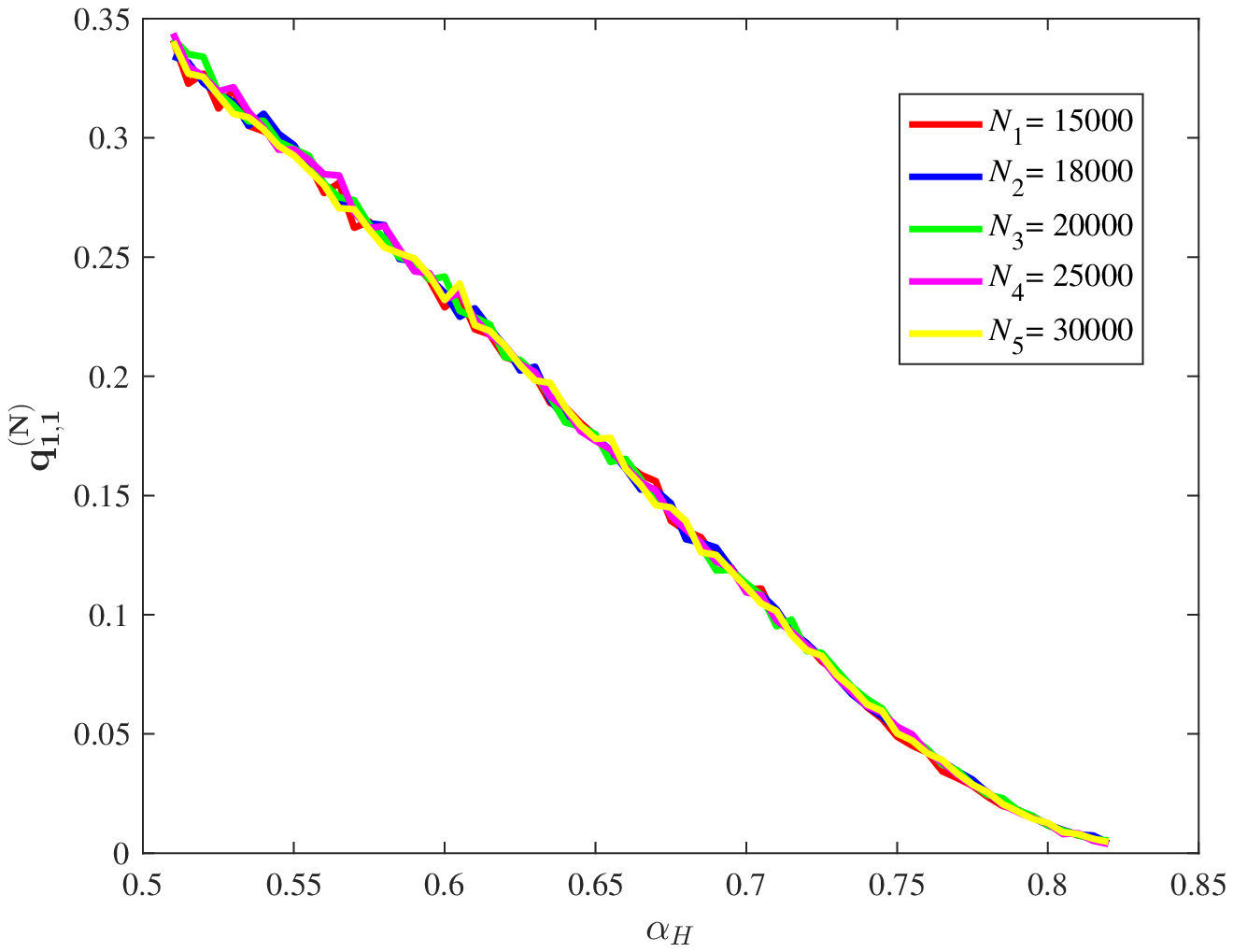}%

\
 \includegraphics[width=1\linewidth]{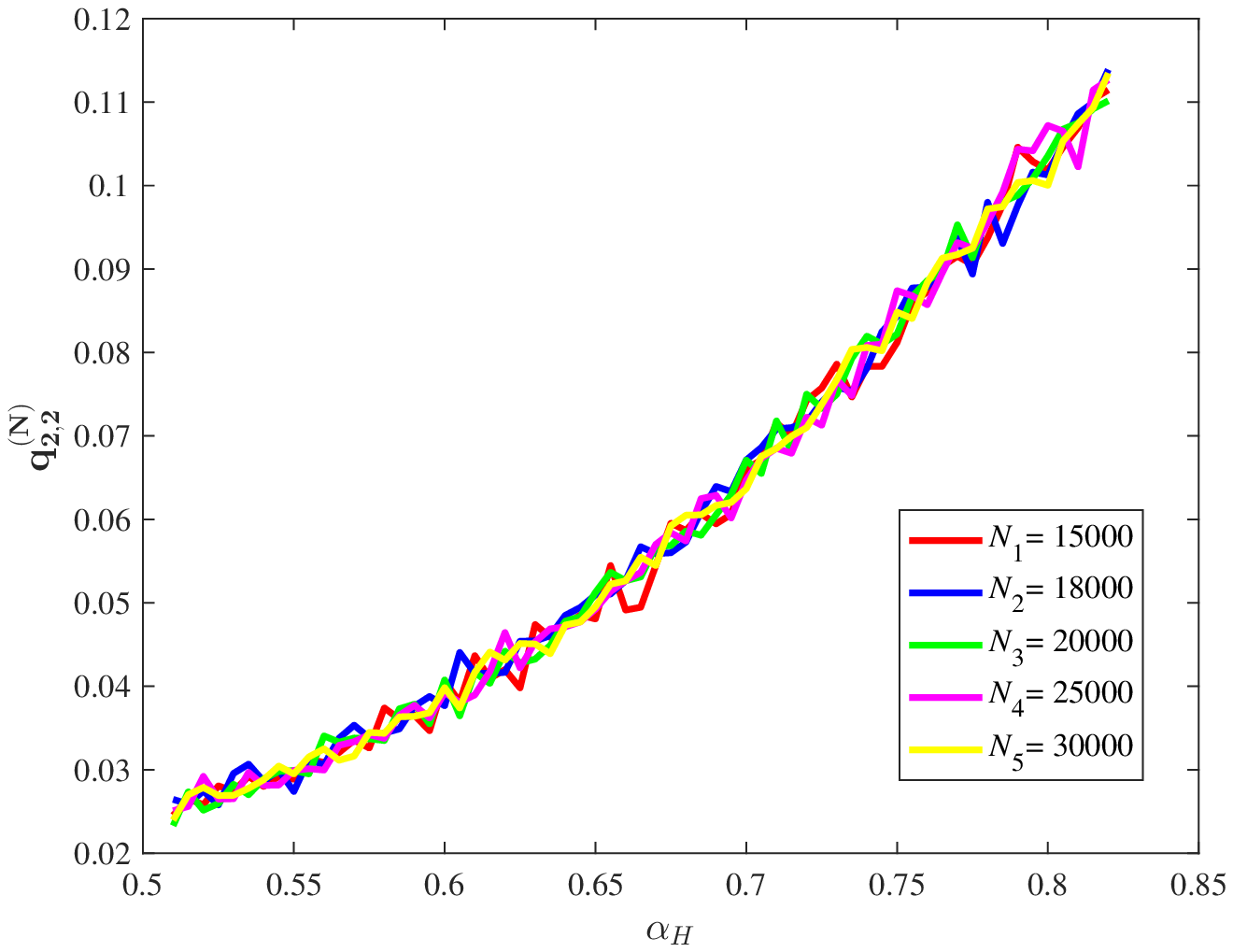}
\end{minipage}}
\subfigure[]{
\begin{minipage}[]{0.4\linewidth}%

\includegraphics[width=1\linewidth]{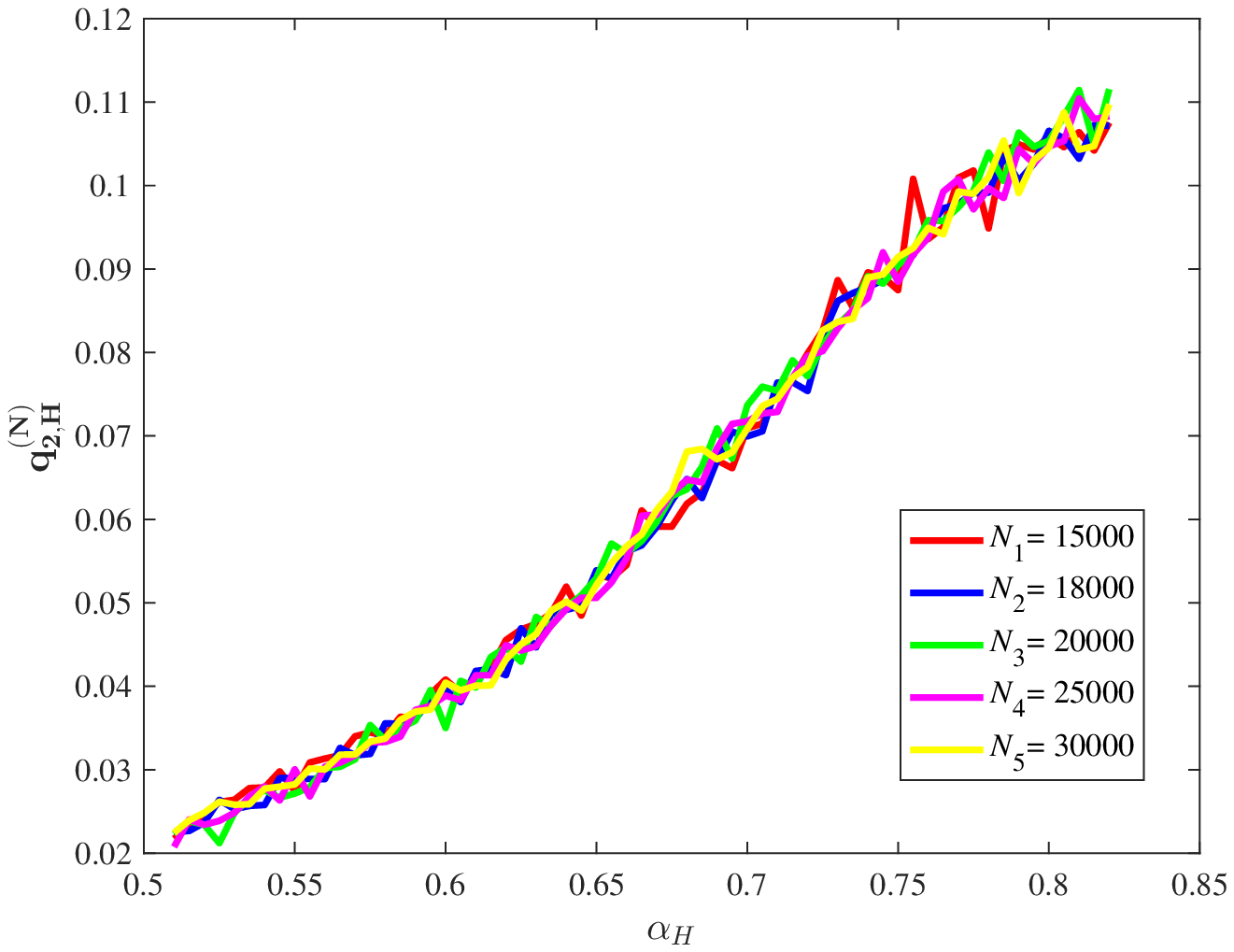}%

\
 \includegraphics[width=1\linewidth]{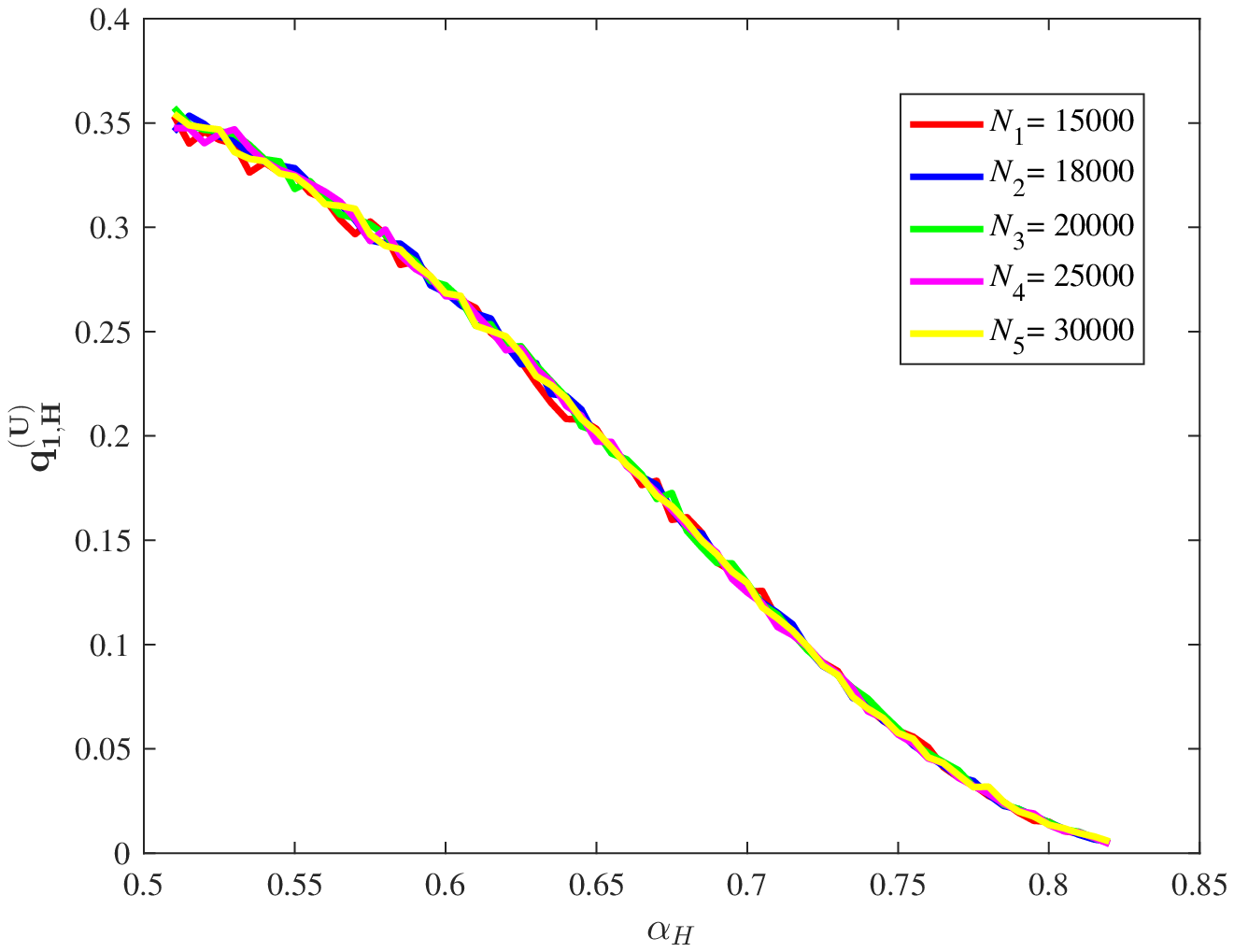}%

\
 \includegraphics[width=1\linewidth]{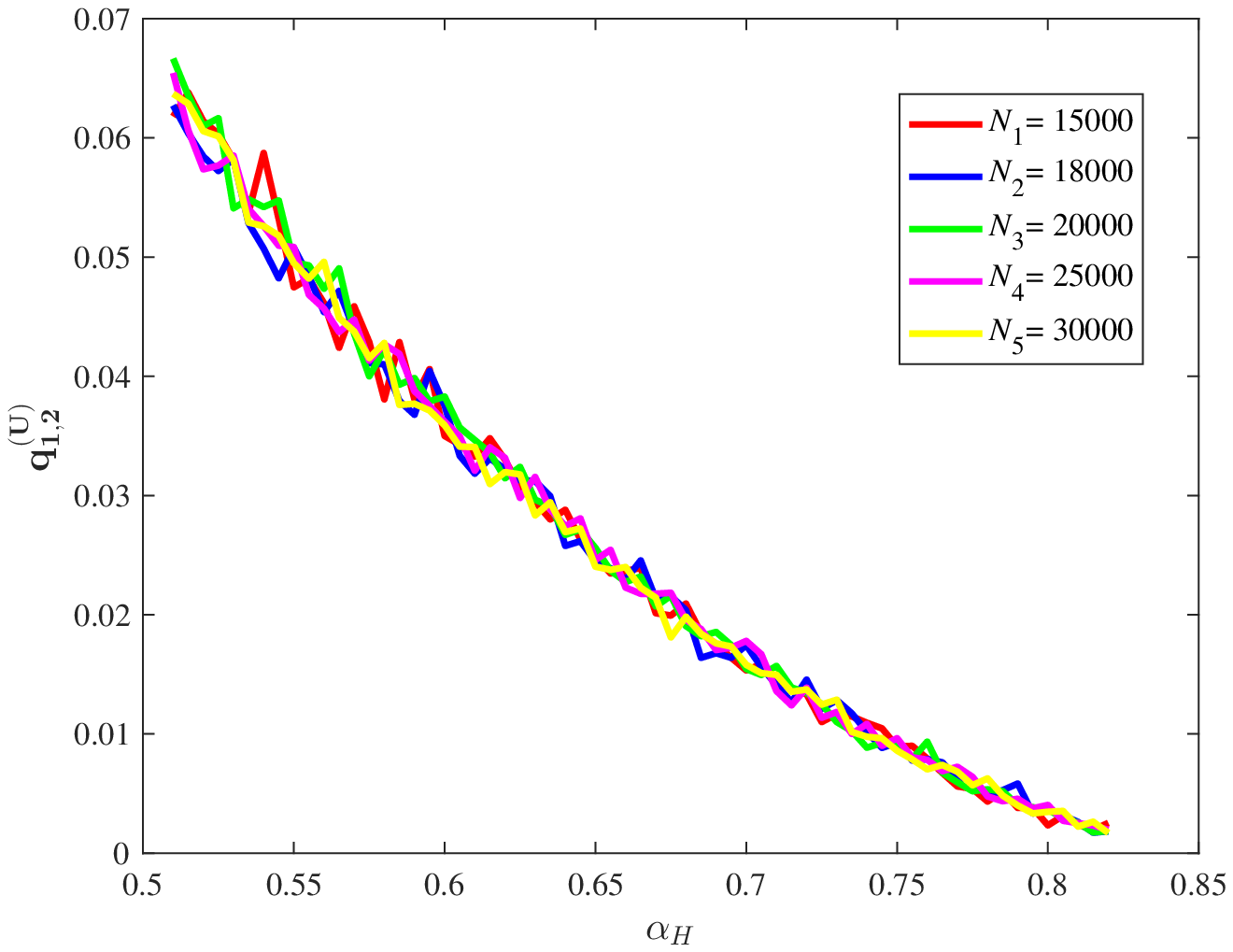}
\end{minipage}}
\caption{Some probabilities that each main chain contains uncle or nephew
blocks. }%
\label{Fig-14}%
\end{figure}

From Figure \ref{Fig-14}, it is\ seen that each of the probabilities is stably
close to a certain value. This shows that the law of large numbers is
successfully applied in our computation. See Theorem 3 for a comparison.

\textbf{(3) Some key ratios vs. $\alpha_{\text{H}}$}

\begin{figure}[ptbh]
\centering          \includegraphics[width=7cm]{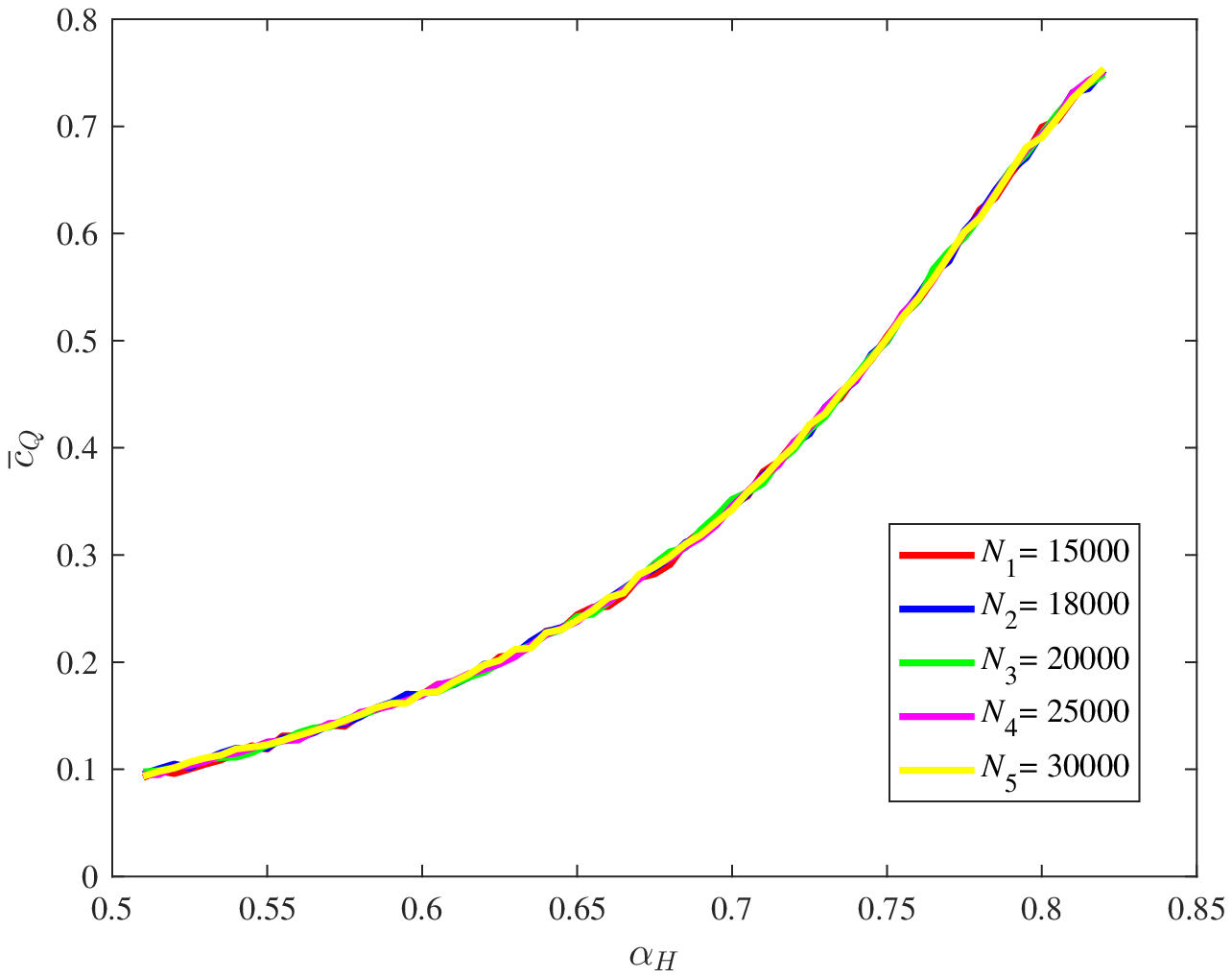}  \centering
\includegraphics[width=7cm]{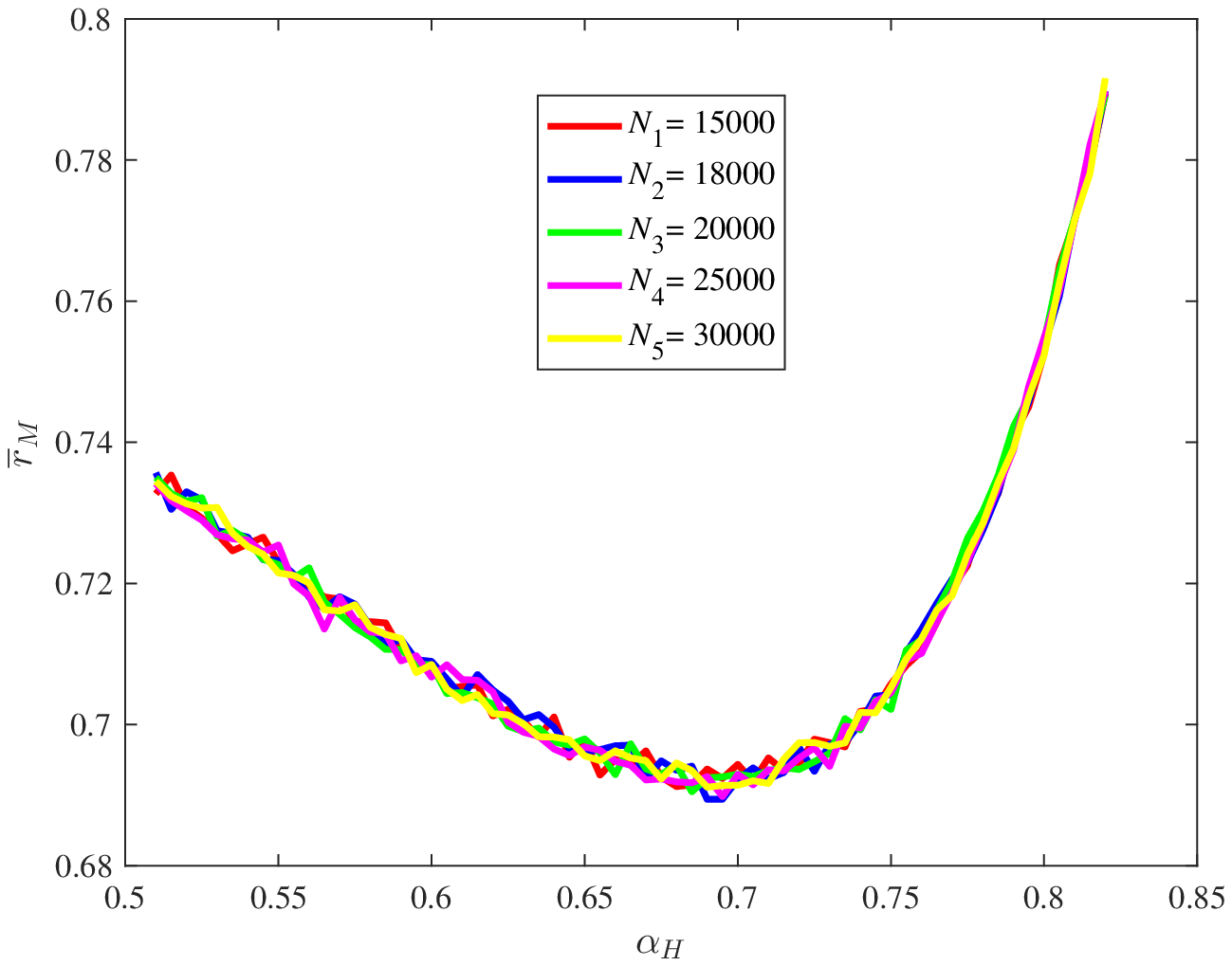}  \centering
\includegraphics[width=7cm]{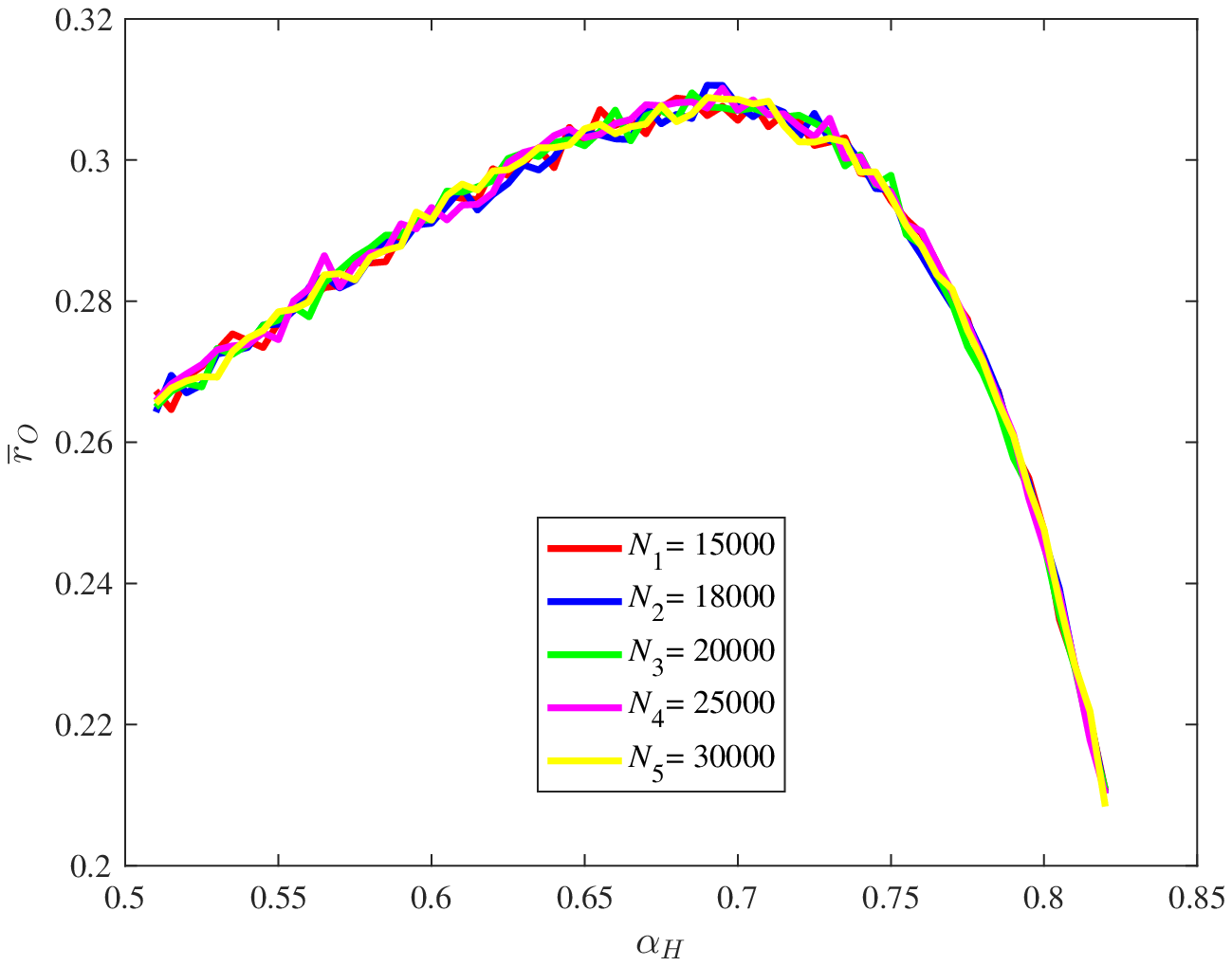}  \centering
\includegraphics[width=7cm]{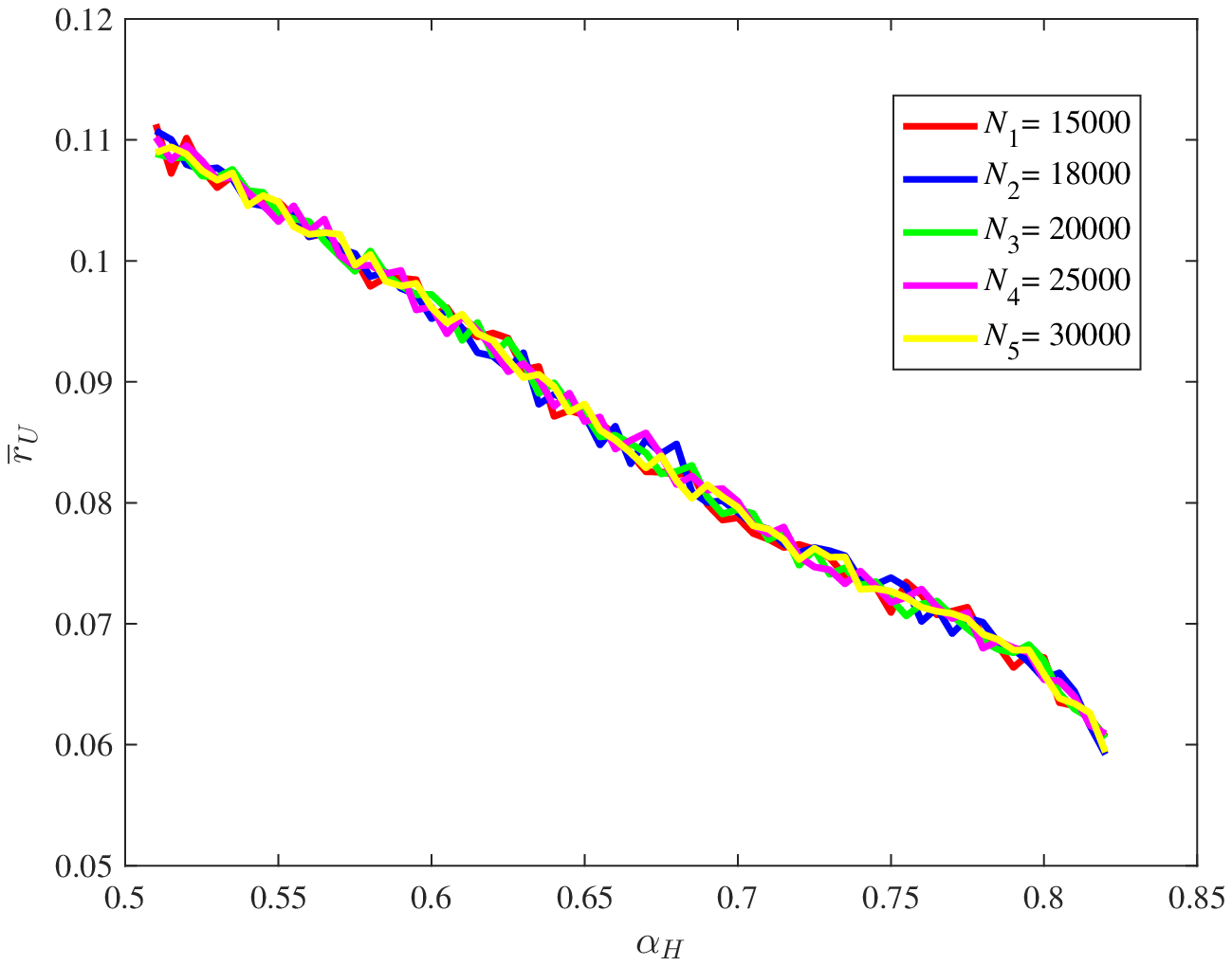}  \centering
\includegraphics[width=7cm]{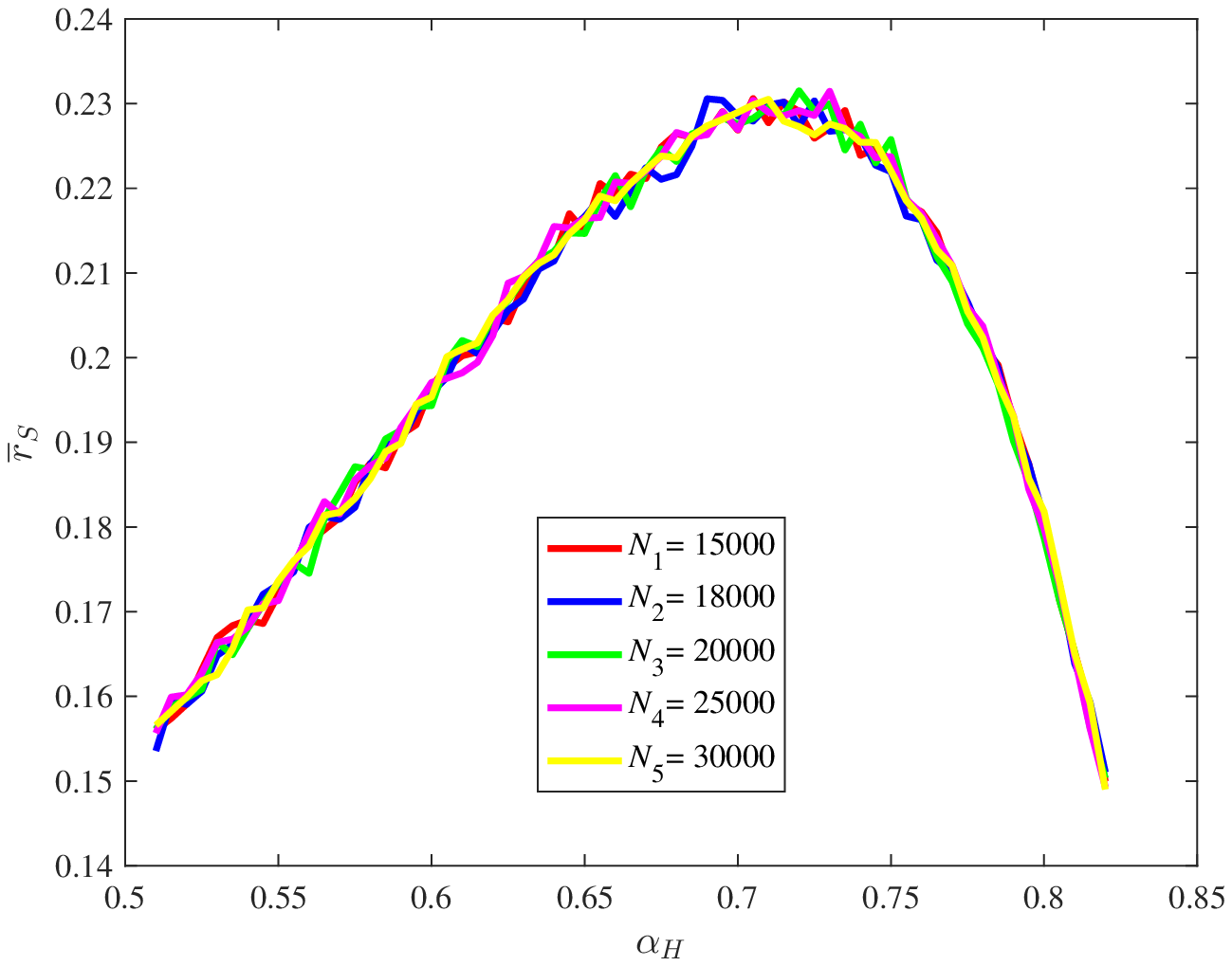}  \caption{The key ratios of
blockchain. }%
\label{Fig-15}%
\end{figure}

From Figure \ref{Fig-15}, it is observed that for some different values of
$N,$ the key ratios $\overline{c}_{\text{Q}}$, $\overline{r}_{\text{M}},$
$\overline{r}_{\text{O}}$, $\overline{r}_{\text{U}}$, and $\overline
{r}_{\text{S}}$ are all approximately stable in our computation by means of
the law of large numbers. Also, the chain quality $\overline{c}_{\text{Q}}$
increases as $\alpha_{\text{H}}$ increases, while the uncle block ratio
$\overline{r}_{\text{U}}$ decreases as the $\alpha_{\text{H}}$ increases. For
the main chain length ratio $\overline{r}_{\text{M}},$ it first decreases and
then increases as the $\alpha_{\text{H}}$ increases. For the stale block ratio
$\overline{r}_{\text{S}}$ (resp. $\overline{r}_{\text{O}}$), it first
increases and then decreases as $\alpha_{\text{H}}$ increases. It shows
from $\overline{r}_{\text{M}},$ $\overline{r}_{\text{O}}$ and $\overline
{r}_{\text{S}}$ that when the mining power of the honest mining pool is close
to $0.7$, the mining competition among the three mining pools is the most
intense, so that a lot of mining resources are wasted.

\textbf{(4) The growth rate of blockchain vs. $\alpha_{\text{H}}$}

\begin{figure}[ptbh]
\centering          \includegraphics[width=10cm]{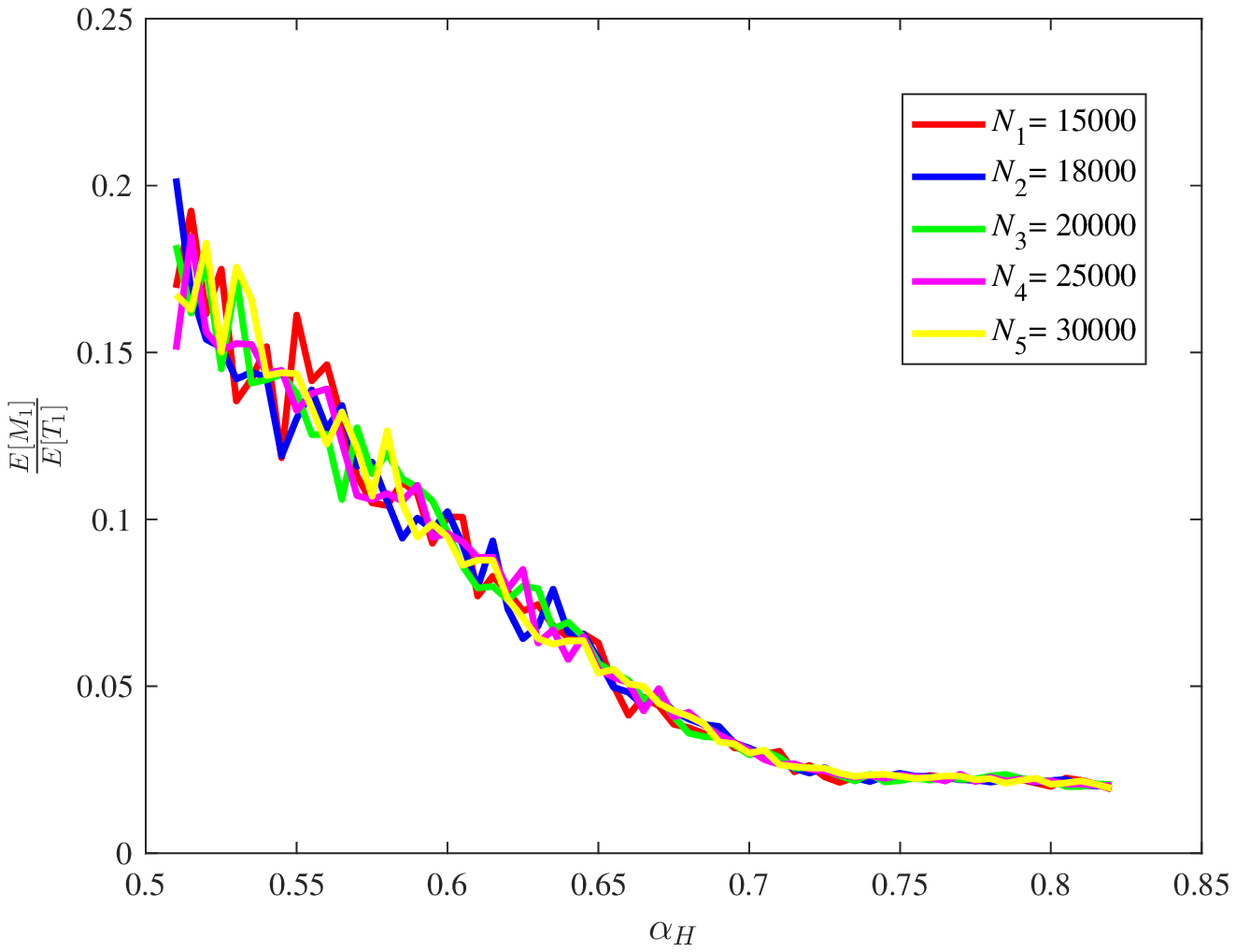}  \caption{The
growth rate of blockchain. }%
\label{Fig-16}%
\end{figure}

From Figure \ref{Fig-16}, we can see that for some different values of
$N$, the growth rate of the blockchain $lim_{t\rightarrow+\infty}M\left(
t\right)  /t=E\left[  M_{1}\right]  /E\left[  T_{1}\right]  $ can be
effectively computed by means of the law of large numbers and the renewal
theorem. In addition, there exists a value $\alpha_{0}\in\left(  0,1\right)  $
such that when $\alpha_{\text{H}}<\alpha_{0}$, the growth rate of blockchain
decreases as $\alpha_{\text{H}}$ increases; while when $\alpha_{\text{H}%
}>\alpha_{0}$, the growth rate of blockchain almost unchanged as $\alpha_{\text{H}}$ increases. It indicates that the growth rate of blockchain
reaches the lowest level once the honest mining pool masters the major mining
power of the entire network. Also, it shows from the Figure \ref{Fig-16} that
the total mining power of the Ethereum system is dispersed into the multiple
mining pools benefits the growth rate of blockchain.

\textbf{(5) The reward of the honest mining pool vs. $\alpha_{\text{H}}$}

\begin{figure}[ptbh]
\centering          \includegraphics[width=10cm]{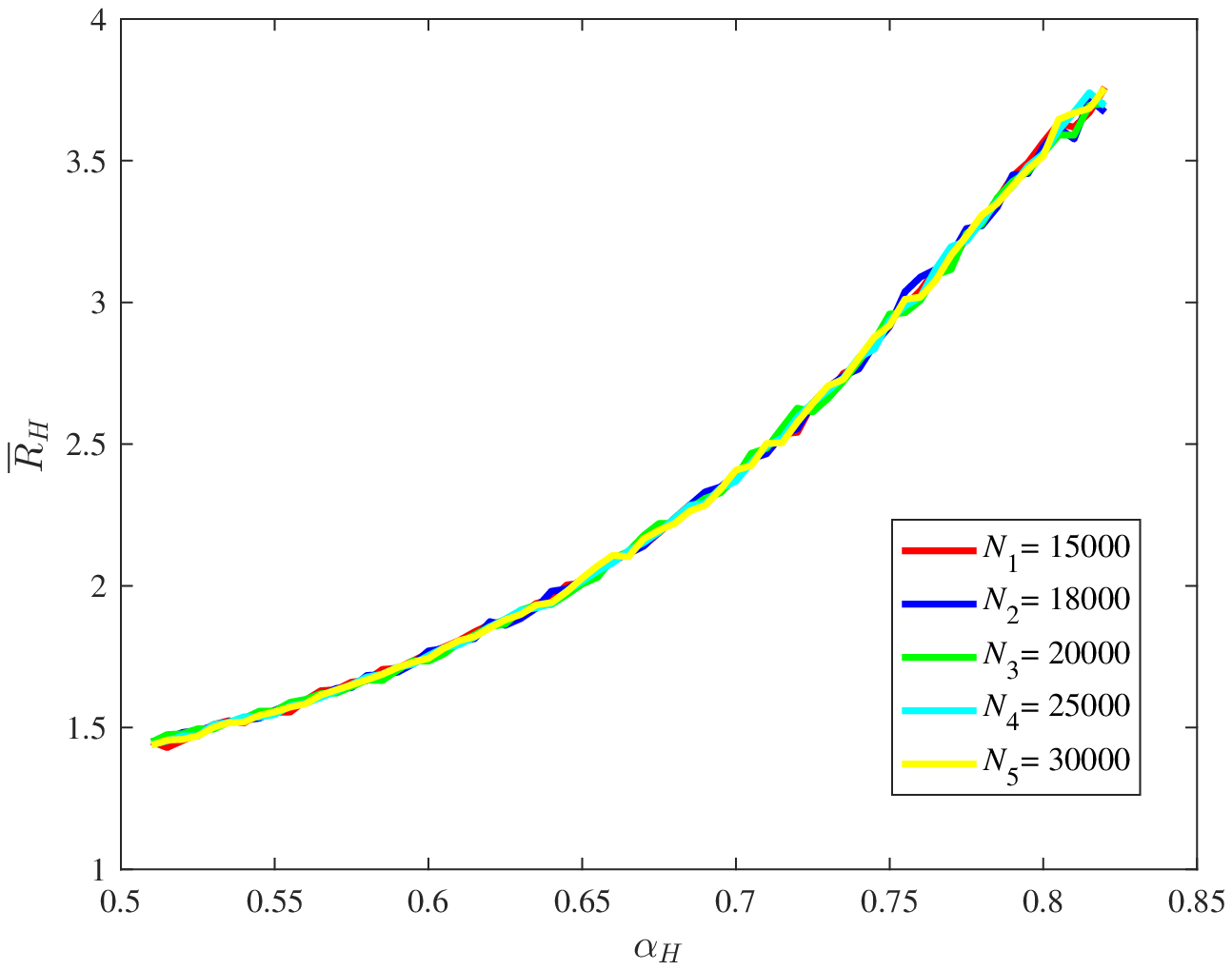}  \caption{The
reward obtained by the honest mining pool. }%
\label{Fig-17}%
\end{figure}

Figure \ref{Fig-17} shows the reward of the honest mining pool. For some
different values of $N$, the reward $\overline{R}_{\text{H}}$ can be
approximately computed by means of the law of large numbers. Also, the reward
$\overline{R}_{\text{H}}$ increases as $\alpha_{\text{H}}$ increases,
which is consistent with the fact that the reward of the honest mining pool is
positively correlated with its mining power.

\textbf{(6) The reward allocation rate of the honest mining pool vs.
$\alpha_{\text{H}}$}

\begin{figure}[ptbh]
\centering             \includegraphics[width=10cm]{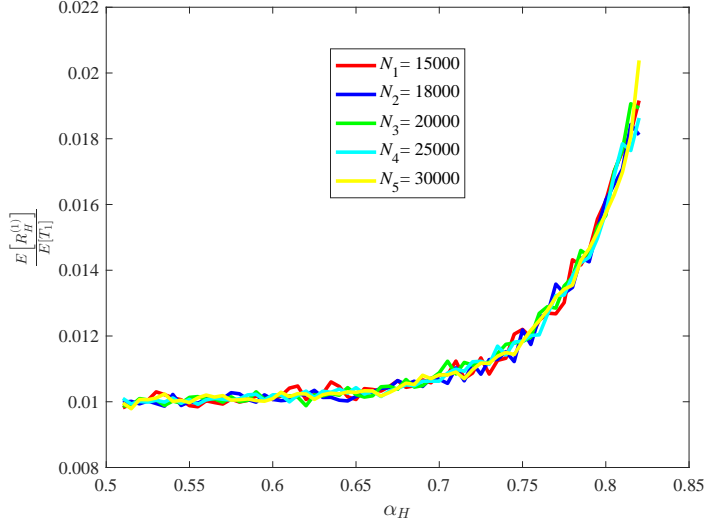}  \caption{The
reward allocation rate of the honest mining pool. }%
\label{Fig-18}%
\end{figure}

Figure \ref{Fig-18} shows the reward allocation rate of the honest mining
pool. For different values of $N$, the reward allocation rate of the honest
mining pool $lim_{t\rightarrow+\infty}R_{\text{H}}\left(  t\right)
/t=E\left[  R_{\text{H}}^{\left(  1\right)  }\right]  /E\left[  T_{1}\right]
$ can be effectively computed by using the law of large numbers and the
renewal theorem. Also, the reward allocation rate increases as
$\alpha_{\text{H}}$ increases, which is also consistent with the fact that the
reward allocation rate of the honest mining pool is positively correlated with
its mining power.

\section{Concluding Remarks}

\label{Sec-11:conclusion}

The growth of the PoW Ethereum system with multiple mining pools has created
the need for not only development of blockchain technology but also setting up
a general mathematical representation of tree and dealing with the
multi-dimensional stochastic systems related to the multiple block branches of
tree. In general, the research on such a tree and associated mathematical
analysis\ is very difficult and challenging. It is worthwhile to note that our
mathematical representation of tree is the first one in the study of
blockchain with multiple mining pools, and it is different from that tree of
the GHOST protocol given in Sompolinsky and Zohar \cite{Som:2013, Som:2015}.

For a blockchain system with two mining pools, Eyal and Sirer \cite{Eya:2014}
found the selfish mining and constructed a simple tree with two block
branches. Following the tree with two block branches, Li et al. \cite{Li:2021}
established the two-dimensional Markov (reward) processes to analyze the
efficiency and benefit of blockchain. However, so far a little research has
worked on the blockchain systems with multiple mining pools although we need to
answer questions such as how to mathematically represent a general tree with
multiple block branches and how to analyze a complicated multi-dimensional
stochastic system running on the general tree. It is obvious that the study of
PoW Ethereum system with multiple mining pools will need to apply the
multi-dimensional stochastic processes on a general tree, even simply, the
fluid and diffusion approximations on a general tree.

In this paper, we described a PoW Ethereum system with multiple mining pools,
which is controlled by the two-block leading competitive criterion proposed in
Li et al. \cite{Li:2021}. Here, a block branch will be generated by only one
mining pool, Thus the mining competition among the multiple mining pools can
generate a general tree with multiple block branches. When observing the
general tree, one of our key findings is to learn that the block
branches of the multiple dishonest mining pools can be forked at any (different)
positions of the block branch of one honest mining pool. Based on this, we can
provide a mathematical representation for the general tree with multiple block
branches. Also, we can easily determine the main chain by means of the
principle of longest chain, e.g., see Li et al. \cite{Li:2021} for the
blockchain with multiple mining pools.

By using the tree representation and observing multiple rounds of mining
competitions, we can provide a block classification of Ethereum: Regular
blocks (i.e., the main chain), orphan blocks, uncle blocks, stale blocks, and
nephew blocks, and set up an approximate computation for the key probabilities
of generating the different types of blocks by applying the law of large
numbers. Based on the key probabilities, together with the tree
representation, we develop an economic framework for computing the rewards
allocated to the multiple mining pools. This is one of our key theoretical
findings in the study of PoW Ethereum system with multiple mining pools.

By applying the renewal reward theorem, we further discuss the growth rate of
blockchain, the reward allocation among the multiple mining pools, and the
reward rates allocated among multiple mining pools, three of which become the
key performance measures of PoW Ethereum system with multiple mining pools.
Furthermore, we use simulation experiments to verify our
theoretical results, and shows that our approximate computation is fast and
effective for dealing with the three performance measures. Therefore, this
paper provides a powerful tool for the performance evaluation of the PoW
Ethereum system with multiple mining pools.

To the best of our knowledge, this paper is the first one to provide the
mathematical representation of general tree, and to analyze the PoW Ethereum
system with multiple mining pools through applying the law of large numbers
and the renewal reward theorem. Therefore, we hope that our methodology and
results given in this paper are applicable to the study of more general PoW
Ethereum system with multiple mining pools. Along the research line, there are
still a number of interesting directions for future research:

\begin{itemize}
\item Setting up a new tree representation for the PoW Ethereum system with
multiple honest mining pools and multiple dishonest mining pools. In the more
complicated case, how to determine the main chain from such a tree? How to
give the performance evaluation of the PoW Ethereum systems?

\item Developing some more effective simulation techniques in the study of PoW
Ethereum system with multiple (honest and dishonest) mining pools through
applying the law of large numbers and the renewal reward theorem.

\item Developing fliud approximation and/or diffusion approximation to analyze
the PoW Ethereum system with multiple (honest and dishonest) mining pools.

\item Providing optimal methods and dynamic control (e.g., Markov decision
processes and stochastic game) in the study of PoW Ethereum system with
multiple (honest and dishonest) mining pools.
\end{itemize}

\section*{Acknowledgements}

Quan-Lin Li was supported by the National Natural Science Foundation of China
under grants No. 71671158 and 71932002.

\end{document}